\documentclass[preprint]{aastex}
\def\la{\lambda}
\def\F{\mathcal{F}}
\def\D{\Delta}

\def\s{\sigma}
\def\ti{\tilde}

\def\integ{\int_{-\infty}^{\infty}}

\usepackage[dvips]{color}
\usepackage{here}
\usepackage{amsmath,amssymb}
\slugcomment{}
\shortauthors{Hirano et al.}
\shorttitle{Improved Modeling of the Rossiter-McLaughlin Effect}
\begin{document}
%%%%%%%%%%%%%%%%%%%%%%%%%%%%%%%%%%%%%%%%%%%%%%%%%%%%%%%%%%%%%%%%%%%%%%
%%%%%%%%%%%%%%%%%%%%%%%%%%%%%%%%%%%%%%%%%%%%%%%%%%%%%%%%%%%%%%%%%%%%%%
\title{Improved Modeling of the Rossiter-McLaughlin Effect for Transiting Exoplanets}
%%%%%%%%%%%%%%%%%%%%%%%%%%%%%%%%%%%%%%%%%%%%%%%%%%%%%%%%%%%%%%%%%%%%%%
%%%%%%%%%%%%%%%%%%%%%%%%%%%%%%%%%%%%%%%%%%%%%%%%%%%%%%%%%%%%%%%%%%%%%%
\author{
Teruyuki Hirano\altaffilmark{1,2}, 
%and TBD
Yasushi Suto\altaffilmark{2,3,4},
 Joshua N.\ Winn\altaffilmark{1},
 Atsushi Taruya\altaffilmark{2,3,5},
 Norio Narita\altaffilmark{6}, 
 Simon Albrecht\altaffilmark{1}, and
 Bun'ei Sato\altaffilmark{7}
} 
%%%%%%%%%%%%%%%%%%%%%%%%%%%%%%%%%%%%%%%%%%%%%%%%%%%%%%%%%%%%%%%%%%%%%%
\altaffiltext{1}{Department of Physics, and Kavli Institute 
for Astrophysics and Space Research, Massachusetts Institute of Technology,
Cambridge, MA 02139}
\altaffiltext{2}{Department of Physics, The University of Tokyo, 
Tokyo 113-0033, Japan}
\altaffiltext{3}{Research Center for the Early Universe, School of Science, 
The University of Tokyo, Tokyo 113-0033, Japan
}
\altaffiltext{4}{Department of Astrophysical Sciences, 
Princeton University, Princeton, NJ 08544}
\altaffiltext{5}
{Institute for the Physics and Mathematics of the Universe (IPMU), 
The University of Tokyo, Chiba 277-8582, Japan}
\altaffiltext{6}{National Astronomical Observatory of Japan, 
2-21-1 Osawa, Mitaka, 
Tokyo, 181-8588, Japan}
\altaffiltext{7}{
Department of Earth and Planetary Sciences, Tokyo Institute of Technology,
2-12-1 Ookayama, Meguro-ku, Tokyo, 152-8551, Japan
}

\email{hirano@utap.phys.s.u-tokyo.ac.jp}
%%%%%%%%%%%%%%%%%%%%%%%%%%%%%%%%%%%%%%%%%%%%%%%%%%%%%%%%%%%%%%%%%%%%%%
\begin{abstract}

  We present an improved formula for the anomalous radial velocity
  of the star during planetary transits due to the Rossiter-McLaughlin
  (RM) effect.  The improvement comes from a more realistic description
  of the stellar absorption line profiles, taking into account stellar
  rotation, macroturbulence, thermal broadening, pressure broadening,
  and instrumental broadening.
  Although the formula is derived for the case in which radial
  velocities are measured by cross-correlation, we show through
  numerical simulations that the formula accurately describes
  the cases where the radial velocities are measured with the iodine
  absorption-cell technique.
  The formula relies on prior knowledge of the parameters describing
  macroturbulence, instrumental broadening and other broadening mechanisms, but
  even 30\% errors in those parameters do not significantly
  change the results in typical circumstances.
  We show that the new analytic formula agrees with previous ones
  that had been computed on a case-by-case basis via numerical simulations.
  Finally, as one application of the new formula, we
  reassess the impact of the differential rotation on the RM velocity anomaly.
  We show that differential rotation of a rapidly rotating star may have
  a significant impact on future RM observations.

\end{abstract}
%%%%%%%%%%%%%%%%%%%%%%%%%%%%%%%%%%%%%%%%%%%%%%%%%%%%%%%%%%%%%%%%%%%%%%
%\keywords{stars: planetary systems: general--stars:rotation--
%techniques: radial velocities--techniques: spectroscopic}
\keywords{planets and satellites: general -- planets and satellites: formation -- stars: rotation --
techniques: radial velocities -- techniques: spectroscopic}
%%%%%%%%%%%%%%%%%%%%%%%%%%%%%%%%%%%%%%%%%%%%%%%%%%%%%%%%%%%%%%%%%%%%%%

\section{Introduction \label{s:intro}}

Transiting exoplanetary systems provide valuable opportunities to
learn about the nature of the exoplanets, their orbits, and their host stars.  In
particular, when we measure the radial velocity (RV) of a star during
a planetary transit, we see an anomalous Doppler shift (beyond the
usual orbital RV) which is called the Rossiter-McLaughlin (hereafter,
RM) effect \citep{Rossiter1924, McLaughlin1924, Hosokawa1953,
  Albrecht2007}.  It arises because a portion of the rotating stellar
disk is blocked by the planet. The partial occultation brings about a
distortion in the spectral lines, which is manifested as an anomalous
RV depending on the position of the planet on the stellar disk
\citep[see, e.g.,][]{Queloz2000, Ohta2005, Winn2005, Narita2007, 
Cameron2010}.  The time variation of the RV anomaly
reveals the (sky-projected) angle $\la$ between the planetary orbital
axis and the stellar spin axis. Measurements of this angle have
proved to be an important observational clue to the origin of
close-in giant exoplanets.

It is widely assumed that close-in gas giants, of which more than 100
are known, formed at a few AU away from their host stars and
subsequently ``migrated'' inward \citep{Lin1996, Lubow2010}.  Many
planetary migration scenarios have been investigated, and some of them
predict small values of $\la\approx 0^\circ$ while others allow larger
spin-orbit misalignment angles \citep[e.g.,][]{Fabrycky2007, Wu2007, 
Nagasawa2008, Chatterjee2008}. The observed distribution of $\la$ and its
dependence on the host star properties (such as masses and ages) 
can be important clues to understand the origin of close-in giant exoplanets 
\citep{Winn2010a, Fabrycky2009, Morton2011}.

Observations of the RM effect have now become almost routine
\citep{Winn2010_alone, Moutou2011}. However,
it is important to remember that the relationship between the observed
RV anomaly, and the position of the planet on the stellar disk, is not
completely straightforward. This is because the RM effect is actually
a spectral distortion, even though it is frequently studied as though
it were a pure Doppler shift.\footnote{An alternative is to model the
  line profiles directly, as has been done by \citet{Albrecht2007} and
  \citet{Cameron2010}, which can be advantageous in some
  circumstances.} Many alternatives have been pursued to calibrate the
relationship between the observed signal and the underlying parameters
of the planet and star. \citet{Ohta2005} and \citet{Gimenez2006}
derived analytic formulas for the RV anomaly based on the computation
of the first moment (the intensity-weighted mean wavelength) of
distorted spectral lines. This approach is simple and convenient
because the computed velocity anomaly does not depend on the intrinsic
shape of spectral lines \citep{Hirano2010}, and has been useful for
quick computations where high accuracy is not essential, and for
gaining insight into the parameter dependence of the RM velocity
anomaly. \citet{Winn2005}, however, noted that the analytic formula by
\citet{Ohta2005} (hereafter, the OTS formula) deviates from the
results based on a more realistic numerical calibration; they
simulated spectra exhibiting the RM effect for many different
positions and sizes of the planet, and then analyzed the mock spectra
with the same data analysis codes that are used routinely to derive
precise RVs with the High Resolution Echelle Spectrometer (HIRES)
installed on the Keck~I telescope. As a result, they showed that the
OTS formula disagrees with the numerical calibration by about 10\% in
terms of the RM amplitude for the case of HD~209458.  
For this reason, subsequent studies
\citep[e.g., ][]{Winn2005, Narita2009a} have relied
upon numerical calibration of the relation between the anomalous RV
and the position and size of the planet, which is done on a
case-by-case basis depending on the stellar parameters.

It would be more convenient to rely on a single analytic formula than
to perform these laborious numerical simulations for each system. An
analytic treatment also provides insight into the reason for the
limitation of the formulas of \citet{Ohta2005} and
\citet{Gimenez2006}. \citet{Hirano2010} took a step in this direction,
pointing out that the discrepancy between the OTS formula and the
simulated results was a consequence of the algorithm used to estimate
the RV anomaly. While the OTS formula was derived by computing the
first moment of the distorted line, in practice the RVs are computed
by cross-correlating an observed spectrum with a template spectrum of
the same star [e.g., for the High Accuracy Radial velocity Planet
Searcher (HARPS); \citet{2009A&A...506..377T}], or by forward-modeling to fit
an observed spectrum with superimposed iodine absorption lines [e.g.,
for the High Dispersion Spectrograph (HDS) on the Subaru telescope and
Keck/HIRES; \citet{Sato2002, Butler1996}]. Using a simplified
description of a single spectral line (a Gaussian function),
\citet{Hirano2010} compared the RV anomalies derived by computing the
first moment and by cross-correlation. They showed that the two
methodologies yield different velocity anomalies, in a manner that
qualitatively explains the previous numerical findings. In particular
they showed that the deviations between the OTS formula and the
results of cross-correlation are larger for more rapidly rotating
stars.

In this work, we take the next step by developing a more realistic
description of stellar line profiles, in order to derive a more
accurate analytic formula. Instead of using a simple Gaussian model
for a spectral line profile, we include realistic kernels for
rotational broadening, macroturbulence and other effects such as
instrumental broadening due to the finite resolution of a
spectrograph.  We test and validate the new analytic formula through
various numerical simulations, and show that it 
is accurate enough for the real data analysis.

This paper is organized as follows. In Section \ref{sec:2}, we derive
the new analytic formula for the RM effect assuming an analytic
function for the stellar line profile.  We present the definitions and
the result there, while the detailed derivation of the main finding
is described in Appendices \ref{sec:app1} and \ref{sec:app2}.  In
order to make sure that the new analytic formula is a good
approximation for the observed velocity anomaly due to the RM effect,
we compare it with numerical simulations using mock transit spectra in
Section \ref{sec:vindication}.  Also, we check on the magnitude of
systematic errors due to imperfect knowledge of the parameters
describing the absorption line profiles for a given star.  
As an application of the new analytic formula, we try to reassess
the impact of stellar differential rotations on the RM velocity anomaly
taking the XO-3 system as a test case in Section \ref{sec:app3}. 
The final section (\S \ref{sec:summary}) is devoted to discussion and summary.

%%%%%%%%%%%%%%%%%%%%%%%%%%%%%%%%%%%%%%%%%%%%%%%%%%%%%%%%%%%%%%%%%%%%%%
\section{Derivation of the New Analytic Formula for the RM Effect\label{sec:2}}

\begin{figure}[H]
\begin{center}
\includegraphics[width=12cm]{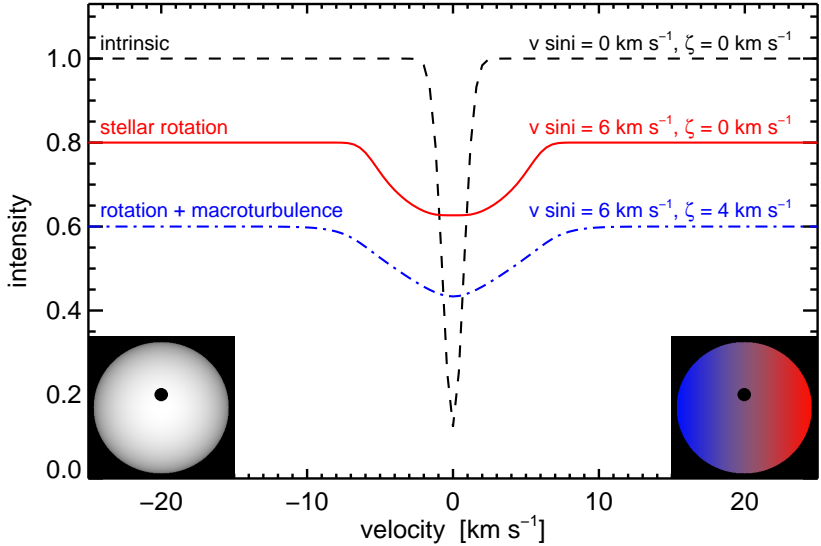} 
\caption{A schematic plot of the line profile during a planetary transit.
Each line has a different broadening kernel.
For visual clarity, the line profiles are vertically shifted by $0.2$
along the intensity axis.
Black line: an intrinsic line profile 
without stellar rotation and macroturbulence, described
as a single Gaussian function with standard deviation $\beta=1$ km s$^{-1}$.
Red line: after convolving with a pure-rotational 
broadening kernel (no macroturbulence), with $v\sin i_s=6$ km s$^{-1}$.
Blue line: after convolving with a rotational-macroturbulent
broadening
kernel
with $v\sin i_s=6$ km s$^{-1}$ and $\zeta=4$ km s$^{-1}$.
For the latter two cases (red and blue lines) the spectral contribution from the portion occulted 
by the planet has been subtracted from the profiles, assuming
a planet with $(R_p/R_s)^2=0.01$.
The line profile with macroturbulence (blue) has elongated wings and the transit signal is nearly invisible.
}\label{fig:lineprofile}
\end{center}
\end{figure}

In this section, we derive the new analytic formula that describes the
velocity anomaly during a transit. We begin with our description of
the stellar absorption lines.  We follow the formulation by
\citet{Hirano2010} but slightly change the basic equations in order
to describe the stellar line profiles more realistically.  Since the
velocity field on the stellar surface is of primary importance, it is
more convenient to express all the functions in terms of velocity
rather than wavelength.  In what follows, the velocity component $v$
indicates the velocity shift relative to the center of an
absorption line. This is related to the wavelength shift $\D\la$ by
the usual formula $\D\la/\la_0=v/c$, where $\la_0$ is the central
wavelength of the absorption line and $c$ is the speed of light.
Following the model of spectral lines by \citet{Gray2005}, we write
a stellar line shape $\F_\mathrm{star}(v)$ as
%%%%%%%%%%%%%%%%%%%%%%%%%%%%
\begin{eqnarray}
\label{eq:fstar}
\F_\mathrm{star}(v)=-S(v)*M(v),
\end{eqnarray}
%%%%%%%%%%%%%%%%%%%%%%%%%%%%
where $S(v)$ is the intrinsic stellar line shape in the absence of stellar rotation and 
macroturbulence (for which we will give an explicit expression later), and 
$M(v)$ is the broadening kernel due to stellar rotation and 
macroturbulence\footnote{We here assume a symmetric line profile and ignore the convective
blueshift (CB) effect, discussed by \citet{Shporer2011}.}. 
The symbol $*$ indicates a convolution between two functions.
Since the continuum level and the normalization 
factor in the spectrum do not affect the result in estimating the velocity 
anomaly during a transit, for convenience we subtract the continuum
level so that
$\F_\mathrm{star}(v)$ becomes zero in the limit of $v\rightarrow
\pm\infty$.
Furthermore we normalize the spectrum so that
%%%%%%%%%%%%%%%%%%%%%%%%%%%%
\begin{eqnarray}
\label{eq:fintegral}
\int_{-\infty}^{\infty}\F_\mathrm{star}(v)dv=-1.
\end{eqnarray}
%%%%%%%%%%%%%%%%%%%%%%%%%%%%
The minus sign indicates that $\F_\mathrm{star}(v)$ describes
an absorption line. The rotational-macroturbulence broadening 
kernel $M(v)$ is calculated by disk-integrating the Doppler-shift component
of the stellar surface due to both stellar rotation and macroturbulence.
We adopt ``the radial-tangential model'' for macroturbulence, for
which the kernel in the absence of rotation is
%%%%%%%%%%%%%%%%%%%%%%%%%%%%
\begin{eqnarray}
\label{eq:macro}
\Theta(v)=\frac{1}{2\sqrt{\pi}} \left[  \frac{1}{\zeta\cos\theta}e^{-\left(\frac{v}{\zeta\cos\theta}\right)^2}
+\frac{1}{\zeta\sin\theta}e^{-\left(\frac{v}{\zeta\sin\theta}\right)^2} 
\right],
\end{eqnarray}
%%%%%%%%%%%%%%%%%%%%%%%%%%%%
where $\zeta$ is the macroturbulent velocity parameter and $\theta$ is the angle between
our line-of-sight and the normal vector to the local stellar surface \citep[][page 433]{Gray2005}.
The angle $\theta$ is related to the coordinate ($x$, $y$) on the stellar disk by
%%%%%%%%%%%%%%%%%%%%%%%%%%%%
\begin{eqnarray}
\label{eq:theta}
\cos\theta=\sqrt{1-\frac{x^2+y^2}{R_s^2}}, ~\sin\theta=\frac{\sqrt{x^2+y^2}}{R_s},
\end{eqnarray}
%%%%%%%%%%%%%%%%%%%%%%%%%%%%
where the $y$-axis is taken to be along the sky projection of the stellar spin axis,
and $R_s$ is the radius of the star.
Assuming a quadratic limb-darkening law, the disk-integrated line broadening function 
due to stellar rotation and macroturbulence is expressed as
%%%%%%%%%%%%%%%%%%%%%%%%%%%%
\begin{eqnarray}
\label{eq:M}
M(v)=\iint_{\mathrm{entire~disk}}\frac{1-u_1(1-\cos\theta)-u_2(1-\cos\theta)^2}{\pi(1-u_1/3-u_2/6)}
~\Theta(v-x\Omega\sin i_s)~\frac{dx~dy}{R_s^2},
\end{eqnarray}
%%%%%%%%%%%%%%%%%%%%%%%%%%%%
where $u_1$ and $u_2$ are the limb-darkening coefficients,
$\Omega$ is the angular spin velocity of the star, and $i_s$
is the inclination angle of the stellar spin axis relative to the line
of sight \citep{Gray2005}. The Doppler shift
$-x\Omega\sin i_s$ in the function $\Theta(v)$ is the consequence of
stellar rotation, neglecting differential rotation.
As \citet{Gray2005} pointed out, the broadening kernel $M(v)$ cannot be expressed
as a convolution of the two different broadening kernels of the stellar rotation and 
the macroturbulence. As we will show, the coupling between rotational broadening 
and macroturbulent broadening
plays an important role in estimating the velocity anomaly due to the RM effect, especially
when the macroturbulent velocity is appreciable when compared to the rotational velocity
of the star (see the difference between line profiles with and without macroturbulence shown
in Figure \ref{fig:lineprofile}). Indeed, this coupling between
rotation and macroturbulence was neglected in the previous
numerical calibrations by \citet{Winn2005} and others.

%%%%%%%%%%%%%%%%%%%
%%%%%%%%%%%%%%%%%%%
Next, we compute the line shape 
during a planetary transit. During a transit, the spectral contribution of the portion blocked 
by the planet is written as
%%%%%%%%%%%%%%%%%%%%%%%%%%%%
\begin{eqnarray}
\label{eq:planet}
\F_\mathrm{planet}(v)=-S(v)*M^\prime(v),
\end{eqnarray}
%%%%%%%%%%%%%%%%%%%%%%%%%%%%
where $M^\prime(v)$ indicates a kernel similar to that given in
Equation~(\ref{eq:M}) but for which the disk-integration should only be performed
over the blocked part of the stellar surface, rather than the entire
stellar disk.
As long as the planet is sufficiently small relative to the star,
the Doppler shift $-x\Omega\sin i_s$ in Equation~(\ref{eq:M}) is nearly 
constant over the integration region.
Thus, if we define $X$ as the $x$-coordinate of the intensity-weighted
center of the eclipsed portion of the star,
we can remove the macroturbulence kernel $\Theta(v)$ from the integral 
and define the following two useful quantities:  
%%%%%%%%%%%%%%%%%%%%%%%%%%%%
\begin{eqnarray}
\label{eq:fandvp}
&&f\equiv \iint_{\mathrm{occulted~portion}}\frac{1-u_1(1-\cos\theta)-u_2(1-\cos\theta)^2}{\pi(1-u_1/3-u_2/6)}
~\frac{dx~dy}{R_s^2},\\
\label{eq:vpdef}
&&v_p\equiv X\Omega \sin i_s,
\end{eqnarray}
%%%%%%%%%%%%%%%%%%%%%%%%%%%%
so that $\F_\mathrm{planet}(v)$ becomes
%%%%%%%%%%%%%%%%%%%%%%%%%%%%
\begin{eqnarray}
\label{eq:planet2}
\F_\mathrm{planet}(v)=-f~S(v)*\Theta(v-v_p).
\end{eqnarray}
%%%%%%%%%%%%%%%%%%%%%%%%%%%%
The first quantity, $f$, is the instantaneous fractional decrease
in flux due to the transit. The second quantity, $v_p$, is the rotational radial velocity
of the occulted portion of the stellar disk, which is occasionally referred to as
the ``subplanet velocity.''
With these definitions the stellar line profile during a transit is expressed as
%%%%%%%%%%%%%%%%%%%%%%%%%%%%
\begin{eqnarray}
\label{eq:transit}
\F_\mathrm{transit}(v)\equiv\F_\mathrm{star}(v)-\F_\mathrm{planet}(v)=-S(v)*M(v)+f S(v)*\Theta(v-v_p).
\end{eqnarray}
%%%%%%%%%%%%%%%%%%%%%%%%%%%%
It should be noted that the
macroturbulent kernel $\Theta(v)$ remains in the modeled
transit line profile. This treatment is necessary since
the two effects of rotational broadening and macroturbulence are coupled with
each other.
In short, the line profile during a transit expressed 
by Equation (\ref{eq:transit}) is different from the line profile
modeled by \citet{Hirano2010} (Eq.[11]) in two senses:
Equation (\ref{eq:transit}) explicitly involves the effect of macroturbulence,
and it is expressed in terms of velocity.

%%%%%%%%%%%%%%%%%%%
\begin{table}[H]
\caption{
Summary of symbols used in this paper.
}\label{table1}
\begin{center}
\begin{tabular}{l|c|c}
\hline\hline
Symbol & Meaning & Typical Range \\\hline
$f$ & \footnotesize{the instantaneous fractional decrease in flux due to the transit (Eq.~[\ref{eq:fandvp}])}
& 0.00 - 0.02\\
$v_p$ & the subplanet velocity  (Eq.~[\ref{eq:vpdef}]) & $\pm v\sin i_s$\\
$\D v$ & the velocity anomaly due to the RM effect  & -\\
$u_1,~u_2$ & \footnotesize{the limb-darkening parameters for the quadratic limb-darkening law} & 0.3 - 0.5\\
$v\sin i_s$ & the stellar spin velocity & -\\
$T_\mathrm{eff}$ & the stellar effective temperature & -\\
$i_s$ & \footnotesize{the inclination of the stellar spin axis measured from our line-of-sight}
& $0^\circ$ - $90^\circ$\\
$l$ & the latitude on the stellar surface & $\pm90^\circ$\\
$R_s$ & the stellar radius & -\\ 
$x,~y$ & the position of the transiting planet on the stellar disk & $\pm R_s$\\
$\alpha$ &  the coefficient of differendtial rotation & $\pm0.02$ \\
$\beta$& the Gaussian dispersion of spectral lines (Eq.~[\ref{eq:beta}]) & 2.5 - 4.5 km s$^{-1}$\\
$\gamma$& the Lorentzian dispersion of spectral lines & 0.5 - 1.5 km s$^{-1}$\\
$\zeta$ & the macroturbulence dispersion  & 2.0 - 6.5 km s$^{-1}$\\
$\theta$ & \footnotesize{the angle between the line-of-sight
and the normal vector to the stellar surface} & $0^\circ$ - $90^\circ$\\
$\la$ & the spin-orbit misalignment angle &  $\pm180^\circ$\\
$\xi$ & the microturbulence dispersion & 0.0 - 2.0 km s$^{-1}$\\
$\sigma$ & the frequency in the Fourier domain & -\\
$\Omega$ & the angular velocity of the stellar spin & -\\
\hline
\end{tabular}
\end{center}
\end{table}
%%%%%%%%%%%%%%%%%%%
Armed with the preceding results, we now express the
velocity anomaly $\D v$ during a transit in terms of the 
fractional flux decrease $f$ and the subplanet velocity $v_p$.
Basically, we follow \citet{Hirano2010}
in order to compute the best-fit value for the anomalous RVs;
they cross-correlated the spectrum during a transit
with a stellar template spectrum, and then calculated the best-fit value for
the velocity anomaly $\D v$ by maximizing the cross-correlation function 
$C(x)$:
%%%%%%%%%%%%%%%%%%%%%%%%%
\begin{eqnarray}
\label{deriva}
&& \frac{dC(x)}{dx}\Big|_{x=\D v} = 0, \\
\label{Cross}
&& C(x) \equiv \integ \F_{\mathrm{star}}(v-x) \F_\mathrm{transit}(v)dv.
\end{eqnarray}
%%%%%%%%%%%%%%%%%%%%%%%%%
To proceed further, we need a specific model for the intrinsic line shape $S(v)$.
We here adopt the Voigt function for $S(v)$:
%%%%%%%%%%%%%%%%%%%%%%%%%
\begin{eqnarray}
\label{voigt}
S(v)=V(v;\beta, \gamma)&\equiv&G(v;\beta)*L(v;\gamma),\\
G(v;\beta)&\equiv& \frac{1}{\beta\sqrt{\pi}}e^{-v^2/\beta^2},\\ 
L(v;\gamma)&\equiv& \frac{1}{\pi}\frac{\gamma}{v^2+\gamma^2},
\end{eqnarray}
%%%%%%%%%%%%%%%%%%%%%%%%%
where $\beta$ is the thermal velocity parameter and $\gamma$ is the
Lorentzian velocity parameter (due to pressure broadening
or natural broadening).  These parameters are
related to individual stellar properties such as the effective
temperature, surface gravity, and the nature of each
absorption line.  Some line profiles of especially strong absorption
lines (such as the Na D lines) are saturated and intrinsically
different from the Voigt function in shape.  However, most of the lines in
the wavelength region used in RV analyses are relatively weak, by
design, and are well approximated by the Voigt function
in the absence of the stellar rotation and macroturbulence.

Substituting Equations (\ref{eq:fstar}) and (\ref{eq:transit}) into
Equations (\ref{deriva}) and (\ref{Cross}), we compute the velocity
anomaly $\D v$ due to the RM effect.  Since further calculations are
mathematically complicated, we describe the detail of the
derivation in Appendix \ref{sec:app1} and write down the result alone:
%%%%%%%%%%%%%%%%%%%%%%%%%%%%
\begin{eqnarray}
\label{eq:result}
\D v\approx -\frac{f}{2\pi}\frac{\displaystyle \int_0^\infty 
\exp(-2\pi^2\beta^2\sigma^2-4\pi \gamma\sigma)
\tilde{M}(\sigma)\tilde{\Theta}(\sigma)\sin(2\pi \sigma v_p)\s d\sigma }
{\displaystyle\int_0^\infty \exp(-2\pi^2\beta^2\sigma^2-4\pi\gamma\sigma)
\tilde{M}(\s)\left\{\tilde{M}(\s)-f\tilde{\Theta}(\s)\cos(2\pi \s v_p)\right\}\s^2d\s},
\end{eqnarray}
%%%%%%%%%%%%%%%%%%%%%%%%%%%%
where
%%%%%%%%%%%%%%%%%%%%%%%%%
\begin{eqnarray}
\tilde{M}(\sigma)&\equiv&\int_{\infty}^{\infty}M(v)e^{-2\pi i\s v}dv \nonumber\\
&=&\int_0^1\frac{1-u_1(1-\sqrt{1-t^2})-u_2(1-\sqrt{1-t^2})^2}{1-u_1/3-u_2/6}\nonumber\\
&&~~~~~~~\times\{e^{-\pi^2\zeta^2\sigma^2(1-t^2)}+e^{-\pi^2\zeta^2\sigma^2 t^2}\}
J_0(2\pi\sigma v\sin i_s t)
tdt, \label{Mfourier}\\
\label{Thetafourier}
\tilde{\Theta}(\sigma)&=&\frac{1}{2}\left[\exp\{-(\pi\zeta\cos\theta)^2\sigma^2\}
+\exp\{-(\pi\zeta\sin\theta)^2\sigma^2\}\right],
\end{eqnarray}
%%%%%%%%%%%%%%%%%%%%%%%%%
where $J_n(x)$ is the Bessel function of the first kind (see also Appendix \ref{sec:app2}). 
Equation (\ref{eq:result}) is the main finding in the present paper
and will be used in the comparison with simulated results.

%%%%%%%%%%%%%%%%%%%%%%%%%%%%%%%%%%%%%%%%%%%%%%%%%%%%%%%%%%%%%%%%%%%%%%
\section{Validity of the Analytic Formula\label{sec:vindication}}
%In this section, we describe the vindication of the analytic formula derived in 
%the previous section.

%%%%%%%%%%%%%%%%%%%%%%%%%%%%%%%%%%%%%%%%%%%%%
\subsection{Comparison with Numerical Simulations for Subaru/HDS\label{sec:tests}}

One of the major differences between the derivation of the analytic
formula (Eq.~[\ref{eq:result}]) and the manner in which RV data are
actually analyzed is that the analytic formula is based on the
cross-correlation method while the data analysis (for Subaru/HDS and
Keck/HIRES at least) adopts the forward modeling method using the
Iodine cell for a precise wavelength calibration.  Another difference
between them is that the analytic formula assumes only one absorption
line in its derivation while the actual stellar spectra have many
lines differing in depth, shape, and so on. Thus, in order to test the
validity of Equation (\ref{eq:result}), we perform the mock data
simulation described below and compare the results with the analytic
formula.

\subsubsection{Mock Data Simulation\label{s:simulation}}

The mock data simulation with the Iodine RV calibration is 
described in detail by \citet{Winn2005} and \citet{Narita2009a}.
In order to generate mock spectra during a transit,
we begin with theoretically synthesized spectra by \citet{Coelho2005},
whose spectral lines are modeled by incorporating thermal broadening
(including microturbulence), and Lorentzian (natural or pressure) broadening. 
Since intrinsic line profiles depend on the stellar type,
we first test the comparison for a G0 star, whose effective 
temperature is 6000~K. 
When we obtain the synthetic spectrum from the synthetic spectrum library, 
we assume a representative value of the surface gravity ($\log g=4.0$) 
for a G0 star with a planet
and the solar abundance for metallicity ([Fe/H]=0.0). 

Using the synthetic spectrum, we generate the mock spectra during a transit by
the following procedure.
\begin{enumerate}

\item We broaden the synthetic spectrum by convolving the
  rotational-macroturbulent kernel $M(v)$ (Eq.~[\ref{eq:M}]), so that
  it represents the template spectrum of the
  actual star. We adopt the limb-darkening parameters $u_1=0.43$ and
  $u_2=0.31$ \citep{Claret2004}, and the macroturbulence parameter
  $\zeta=4.3$ km s$^{-1}$ \citep{Valenti2005}.  We try three different
  values for the rotational velocity of the star: $v\sin i_s=2.5$ km
  s$^{-1}$, 5.0 km s$^{-1}$, and 7.5 km s$^{-1}$.

\item We Doppler-shift the original unbroadened spectrum by $v_p$,
  multiply by $f$, and then convolve the spectrum with the
  macroturbulence kernel $\Theta(v)$ (Eq.~[\ref{eq:macro}]), so that
  the resultant spectrum represents the spectral contribution from the
  portion occulted by the transiting planet.  For each of the three
  values of $v\sin i_s$, we consider $f=0.004$, 0.008, 0.012, 0.016, and
  0.020.  We then assign 21 different values to $v_p$ evenly spaced
  from $-v\sin i_s$ to $+v\sin i_s$, yielding in total 105 different
  points in the $(f, v_p)$ grid for each value of $v\sin i_s$.  The
  macroturbulent broadening depends not only on the $x$-coordinate but
  also on the $y$-coordinate on the stellar disk (see
  Eqs.~[\ref{eq:macro}] and [\ref{eq:theta}]).  For simplicity,
  however, we assume $y=0$ when we generate the mock transit spectra.

\item We generate the mock transit spectra by subtracting the
  Doppler-shifted and scaled spectra in the second step from the
  broadened spectra created in the first step.

\item We multiply the mock transit spectra by the iodine transmission
  spectrum used for calibration, and convolve the Star+I$_2$ spectrum
  with the representative instrumental profile of Subaru/HDS for the
  case of the slit width being 0.8$^{\prime\prime}$, corresponding to
  the spectral resolution of $R\sim 45000$ \footnote{The choices for
    the instrumental profiles used in the mock simulation are
    explained later.}.

\end{enumerate}

We then take these 105 mock spectra for each of three values of $v\sin
i_s$, and use them as inputs to the RV analysis routine for
Subaru/HDS. The RV analysis using the iodine cell is described in
detail by \citet{Sato2002} (for Subaru/HDS) and \citet{Butler1996}
(for Keck/HIRES).  For each point of the $(f, v_p)$ grid, the RV
analysis routine outputs a velocity anomaly $\D v$ due to the RM
effect.

\subsubsection{Results for a G0 Star\label{sec:resultG0}}

\begin{figure}[H]
\begin{center}
\vspace{-1.5cm}
\includegraphics[width=9.5cm,clip]{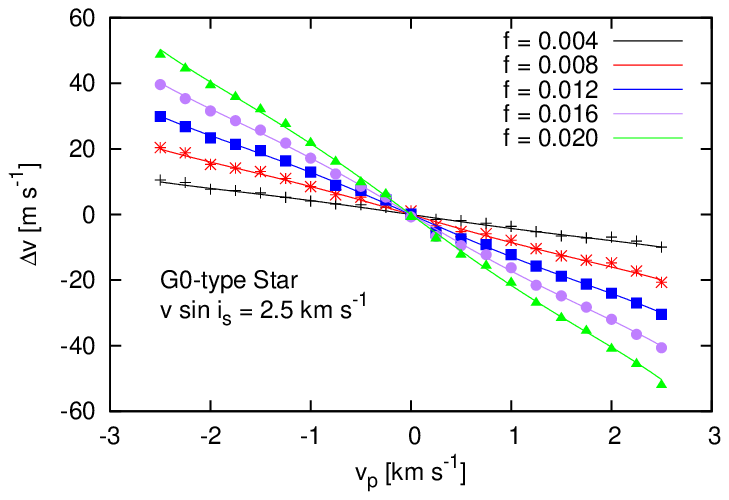} 
\includegraphics[width=9.5cm,clip]{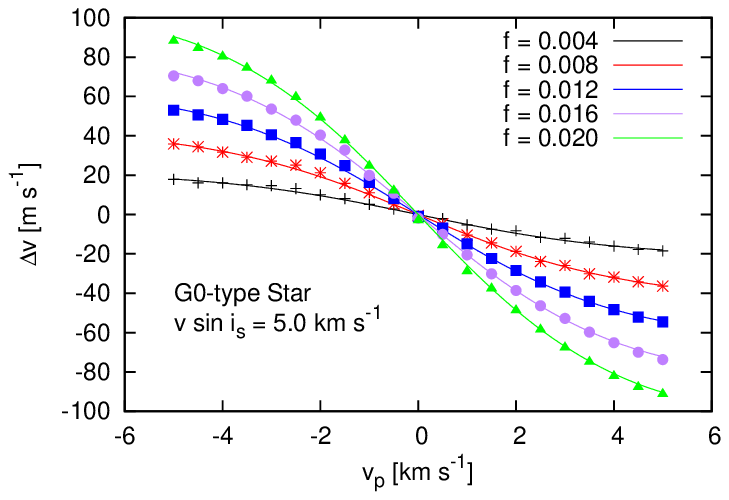} 
\includegraphics[width=9.5cm,clip]{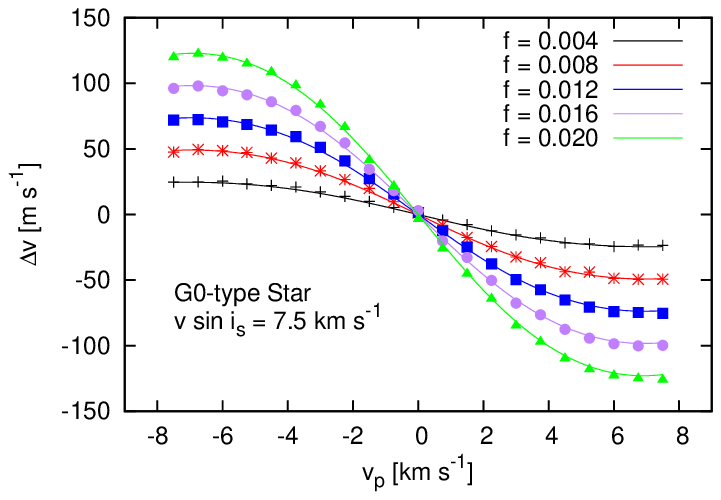} 
\caption{
Simulated velocity anomalies due to the RM effect v.s. the analytic 
formula (Eq. [\ref{eq:result}]) as a function of the subplanet velocity $v_p$
in the case of a G0-type star: $v\sin i_s=2.5$ km s$^{-1}$ (top), 5.0 km s$^{-1}$ (middle), 
and 7.5 km s$^{-1}$ (bottom). In each panel,
symbols in black, red, blue, purple, and green indicate the simulated
data points for $f=0.004$, 0.008, 0.012, 0.016, and 0.020, respectively.
Analytic curves based on Equation (\ref{eq:result}) are plotted 
with the solid lines in the same color as the symbols.
}\label{fig:G0mock}
\end{center}
\end{figure}

We compare the results $(f, v_p, \D v)$ based on the mock data
simulation with the analytic expression (\ref{eq:result}).  Figure
\ref{fig:G0mock} shows the comparison between the simulated data
points $(f, v_p, \D v)$ and the analytic formula
(Eqn.~[\ref{eq:result}]). The three different panels are for the three
different values of $v\sin i_s$: (top) 2.5 km~s$^{-1}$, (middle) 5.0
km~s$^{-1}$, and (bottom) 7.5 km~s$^{-1}$. For each of the five values
of $f$, the data points indicated by color symbols show the simulated
results; $f=0.004$ in black, $f=0.008$ in red, $f=0.012$ in blue, $f=0.016$ in
purple, and $f=0.020$ in green.

In order to draw the analytic curves, we need to choose values of $\beta$ and
$\gamma$. Although they are related to intrinsic stellar line
profiles, each spectral line has its own values for $\beta$ and
$\gamma$, so we need to know the ``effective'' values of those line
parameters in order to make a comparison between the analytic formula
and the simulated results.  For the purpose, we employ an
autocorrelation method. By autocorrelating the synthetic spectrum
for a G0-type star, we obtain an effective line profile of the
spectrum (see Appendix \ref{sec:app2.5} for details).  Since there is a
strong degeneracy between the two stellar line parameters $\beta$ and
$\gamma$, we fix $\beta$ based on a simple physical principle and
estimate the Lorentzian dispersion $\gamma$ from the effective line
profile.  In principle, the intrinsic Gaussian dispersion $\beta_0$ in
each spectral line is determined by the effective temperature
$T_\mathrm{eff}$ of the star and the ``the micro-turbulence"
dispersion $\xi$ as
%%%%%%%%%%%%%%%%%%%%%%%%%%%%
\begin{eqnarray}
\label{eq:beta0}
\beta_0=\sqrt{\frac{2k_\mathrm{B} T_\mathrm{eff}}{\mu}+\xi^2},
\end{eqnarray}
%%%%%%%%%%%%%%%%%%%%%%%%%%%%
where $k_\mathrm{B}$ and $\mu$ are the Boltzmann constant and the mass of the 
atom (or molecule) in question, respectively \citep{Gray2005}.
In addition, as we will show later, the simulated velocity anomalies also depend
on the instrumental profile which we assume in generating mock spectra.
Therefore, we here adopt an \textit{ad hoc} assumption that 
the Gaussian width parameter $\beta$ in Equation (\ref{eq:result}) depends 
also on the width of the instrumental profile so that $\beta$ is expressed as
%%%%%%%%%%%%%%%%%%%%%%%%%%%%
\begin{eqnarray}
\label{eq:beta}
\beta=\sqrt{\beta_0^2+\beta_\mathrm{IP}^2}=
\sqrt{\frac{2k_B T_\mathrm{eff}}{\mu}+\xi^2+\beta_\mathrm{IP}^2},
\end{eqnarray}
%%%%%%%%%%%%%%%%%%%%%%%%%%%%
where $\beta_\mathrm{IP}$ is the dispersion of Gaussian which best-fits
the representative instrumental profile adopted in observations.

Substituting $T_\mathrm{eff}=6000~\mathrm{K}=0.517$ eV, 
$\mu=52$ GeV/c$^{2}$ (the mass of an iron 
atom\footnote{Iron lines are most numerous in the wavelength range used for the RV analysis
with the iodine cell.}), $\xi=1.0$ km s$^{-1}$ \citep{Coelho2005}, and 
$\beta_\mathrm{IP}=3.6$ km s$^{-1}$ (the dispersion of the instrumental 
profile of Subaru/HDS we assumed in generating mock transit spectra,
corresponding to the spectral resolution of $R\sim 45000$), 
we obtain $\beta_0=1.7$ km s$^{-1}$ and 
$\beta=4.0$ km s$^{-1}$. Fitting the effective line profile with 
the Voigt function assuming $\beta_0=1.7$ km s$^{-1}$, 
we obtain $\gamma=0.9$ km s$^{-1}$ (Appendix \ref{sec:app2.5}).

In each panel (different $v\sin i_s$) of Figure \ref{fig:G0mock}, 
the analytic curves based on Equation (\ref{eq:result}) are 
shown in the same colors as the simulated results indicated by symbols for
each value of $f$. We adopt $(\beta,\gamma)$=(4.0 km s$^{-1}$, 0.9 km s$^{-1}$).
The other parameters ($u_1, u_2, v\sin i_s, \zeta$) in Equation (\ref{eq:result})
are fixed at the same values used to make mock transit spectra.
%%%%%%%%%%%%%%%%%%%
%%%%%%%%%%%%%%%%%%%

For all of the three different rotational velocities, the analytic
curves are in very good agreement with the simulated results. As long
as $v\sin i_s$ is small, the velocity anomaly curves are nearly linear
and can be approximated as $\D v \approx -f v_p$. As \citet{Winn2005}
and other authors have pointed out, however, they are significantly
curved for larger values of $v\sin i_s$.

\subsubsection{The Impact of Instrumental Profiles\label{sec:IP}}

In the forward modeling method using the iodine cell, radial
velocities $v_\mathrm{RV}$ of a star are computed by modeling each
observed spectrum $I_\mathrm{obs}(\la)$ with the following equation:
%%%%%%%%%%%%%%%%%%%%%%%%%%%%
\begin{eqnarray}
\label{eq:rv_computation}
I_\mathrm{obs}(\la)=k\left[A(\la)T\left(\la(1-v_\mathrm{RV}/c)\right)\right]*\mathrm{IP},
\end{eqnarray}
%%%%%%%%%%%%%%%%%%%%%%%%%%%%
where $A(\la)$ and $T(\la)$ are the transmission spectrum of the
Iodine cell and the template spectrum of the same star, respectively
\citep{Sato2002, Butler1996}.  Outside of a transit, the intrinsic
stellar line profile of each spectrum is supposed to be the same as
that of the template spectrum, and therefore $v_\mathrm{RV}$ is not
affected by the instrumental profile which may be variable during an
observation.  During a transit, however, the intrinsic stellar line
profile is distorted due to the partial occultation, and it is not
clear if the IP affects the radial velocity $v_\mathrm{RV}$ including
the RM velocity anomaly $\D v$.  Thus, in order to see if the
instrumental profile used in the mock data analysis affects the RM
velocity anomaly, we repeated the same mock data analysis described
above, but with a different instrumental profile.  In the simulation
above, we have fixed the instrumental profile so that it corresponds
to the spectral resolution of $R = 45000$ for
Subaru/HDS.  This time, we adopt the instrumental profile with
the spectral resolution of $R = 90000$. In this case, the dispersion of 
Gaussian fits the instrumental profile is approximately
$\beta_\mathrm{IP}=1.9$ km s$^{-1}$.
After generating many mock spectra for $v\sin i_s=2.5$, 5.0, and 7.5 km s$^{-1}$
as in Section \ref{s:simulation},
we analyzed them with the usual RV routine to obtain the velocity anomaly $\D v$.

In order to quantify the differences between the two cases of the
instrument profile, we compute the following statistics for each case
of the instrumental profiles:
%%%%%%%%%%%%%%%%%%%%%%%%%%%%
\begin{eqnarray}
\label{eq:statistics}
D(\beta,\gamma)\equiv \sqrt{\frac{1}{315}\sum_{f,v_p,v\sin i_s}^{315}\Big|
\frac{\D v_\mathrm{sim}(f,v_p)-\D v_\mathrm{ana}(f, v_p,\beta,\gamma)}
{f\cdot v\sin i_s}\Big|^2},
\end{eqnarray}
%%%%%%%%%%%%%%%%%%%%%%%%%%%%
where $\D v_\mathrm{sim}(f,v_p)$ is the simulated velocity anomaly for
each point of the $(f, v_p)$ grid, and $\D v_\mathrm{ana}(f, v_p,
\beta,\gamma)$ is the value computed by Equation~(\ref{eq:result}) as
a function of $f$, $v_p$, $\beta$, and $\gamma$.  
When we compute $\D v_\mathrm{ana}(f, v_p,\beta,\gamma)$, 
all the parameters other than $\beta$ and $\gamma$ are fixed at exactly 
the same values as are adopted in the numerical simulation (such as $\zeta$ 
and $u_1, u_2$). 
The summation in the
above expression is performed over the $315~(=105\times 3)$ data
points on the $(f, v_p)$ grid with three different values of $v\sin
i_s$.  The statistics, $D(\beta,\gamma)$, indicates the degree of
agreement between the analytic formula and the simulated results in
terms of the representative velocity anomaly $fv\sin i_s$.  For each
of the instrumental profiles ($R=45000$ and $R=90000$), we
compute $D(\beta,\gamma)$, for $2.0~
\mathrm{km~s}^{-1}\leq \beta\leq5.0~ \mathrm{km~s}^{-1}$ and $0.0
~\mathrm{km~s}^{-1}\leq \gamma\leq3.0~ \mathrm{km~s}^{-1}$.

%%%%%%%%%%%%%%%%%%%
\begin{figure}[H]
\begin{center}
\includegraphics[width=11cm,clip]{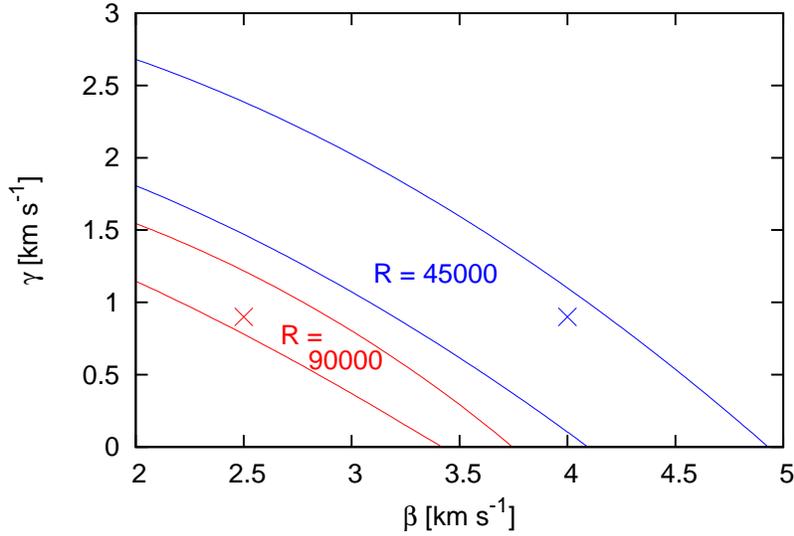} 
\caption{
The contour of $D(\beta,\gamma)$ for a G0-type star.
The regions surrounded by the two blue and the two red curves are the
best-fit regions of the analytic formula,
where $D(\beta,\gamma)\leq0.005$ for $R=45000$  and $R=90000$,
respectively. The values of $(\beta,\gamma)$ estimated by
the line analysis are shown by the blue and red crosses
for $R=45000$  and $R=90000$,
respectively.
}\label{contourG0.8}
\end{center}
\end{figure}
%%%%%%%%%%%%%%%%%%%
As a result of computing $D(\beta,\gamma)$ in the $(\beta,\gamma)$
grid, we find the lowest value of $D(\beta,\gamma)$ is approximately
0.0045 for both of the two different instrumental profiles,
representing very good agreement.  For instance, the dispersion of the
simulated velocity anomalies around the analytic formula is expected
to be less than $\sim 0.25$~m~s$^{-1}$ in case of $f\approx 0.01$ and
$v\sin i_s\approx 5.0$~km~s$^{-1}$.  This deviation is much smaller than
the usual precision with which RVs can be measured.

In Figure \ref{contourG0.8}, we show contours of the goodness-of-fit
statistic $D(\beta,\gamma)$.
The regions between the two blue curves and two red curves are
where $D(\beta,\gamma)\leq 0.005$ for $R=45000$ and $R=90000$,
respectively. 
Since the Gaussian and Lorentzian dispersions $\beta$ and $\gamma$ strongly
degenerate, the values of $\beta$ and $\gamma$ that fit the simulated
results well are expected to be located in extended areas.
Thus, we show the ``best-fit regions'' where the analytic formula
agrees well with the simulated results. 
From Figure \ref{contourG0.8}, it is obvious that the best-fit region 
is shifted toward smaller values of $(\beta,\gamma)$ when
we adopt the higher spectral resolution.
This result implies that the instrumental profile in 
spectroscopic observations does affect the velocity anomaly due to the RM effect.
We interpret the results as follows; in the forward-modeling fitting procedure,
the instrumental profile (a finite spectral resolution) plays a similar role 
as the other physical broadening mechanisms for spectral lines.
For reference, we show $(\beta,\gamma)$=(4.0 km s$^{-1}$, 0.9 km s$^{-1}$) 
by the blue cross, which are the intrinsic line parameters for 
$R=45000$ estimated 
by the spectral line analysis in Appendix \ref{sec:app2.5}.
Also, substituting $\beta_\mathrm{IP}=1.9$ km s$^{-1}$ into Equation (\ref{eq:beta})
for the case of $R\sim 90000$,
we obtain $\beta=2.5$ km s$^{-1}$, which, along with $\gamma=0.9$ km s$^{-1}$
(the same value as used in the comparison for $R= 45000$),
is shown in Figure \ref{contourG0.8} by the red cross.
It should be emphasized that the values of $(\beta, \gamma)$
shown by the blue and red crosses are estimated in a way independent
of the mock data simulation (Appendix \ref{sec:app2.5}) but
are consistent with the regions where the analytic formula 
best agrees with the simulated results,
for the two different instrumental profiles.

In summary,
as a result of trying two different cases of the spectral resolution ($R\sim 45000$ and 90000), 
we have shown that the RM velocity anomaly is actually affected by the 
specific choice of the instrumental profile. Since instrumental profiles are often
approximated as Gaussian, we incorporate its impact on the line profile into the intrinsic 
Gaussian dispersion $\beta_0$ by adding the instrumental broadening $\beta_\mathrm{IP}$
in quadrature, which justifies the treatment shown in Equation (\ref{eq:beta}).

\subsubsection{Results for Other Spectral Types of Stars}

So far, we have considered a G0 star. In order to make sure
that our analytic formula is applicable for a variety of different stellar
types, we consider stars with the effective
temperatures of $T_\mathrm{eff}=6500$~K (an F5 star) and 5000~K
(a K0 star) using the theoretically synthesized spectra, just as
we did for a G0 star.

\begin{description}
\item[F5 star] 
In generating mock spectra for an F-type star ($T_\mathrm{eff}=6500$ K), 
we broaden the synthetic spectrum of an F5 star
\citep{Coelho2005} assuming the rotational velocity of
$v\sin i_s=5.0$, 10, and 15 km s$^{-1}$.
The other adopted parameters include $u_1=0.32$ and $u_2=0.36$ 
for the limb darkening parameters \citep{Claret2004}, and
$\zeta=6.2$~km~s$^{-1}$ for the macroturbulent velocity parameter \citep{Gray2005}.
Except for those parameters, we perform exactly the same simulation described 
in Section \ref{s:simulation}. 
The simulated results for $v\sin i_s=10$ km s$^{-1}$ are shown in 
the upper panel of Figure \ref{fig:F5mock},
along with the analytic formula (Eq.~[\ref{eq:result}]) assuming 
$(\beta, \gamma)=(4.0$ km s$^{-1}$, 0.9 km s$^{-1}$).
These Gaussian and Lorentzian dispersions are also estimated 
by the combination of Equation~(\ref{eq:beta})
adopting $T_\mathrm{eff}=6500$~K, 
and the spectral line analysis using the auto-correlation method.
Again, the analytic curves 
well describe the behavior of the simulated results. They are also in good 
agreement with each other for $v\sin i_s=5.0$ km s$^{-1}$ and 15 km s$^{-1}$,
although for brevity we do not show all of those results here.
%%%%%%%%%%%%%%%%%%%
\begin{figure}[H]
\begin{center}
\includegraphics[width=9.5cm,clip]{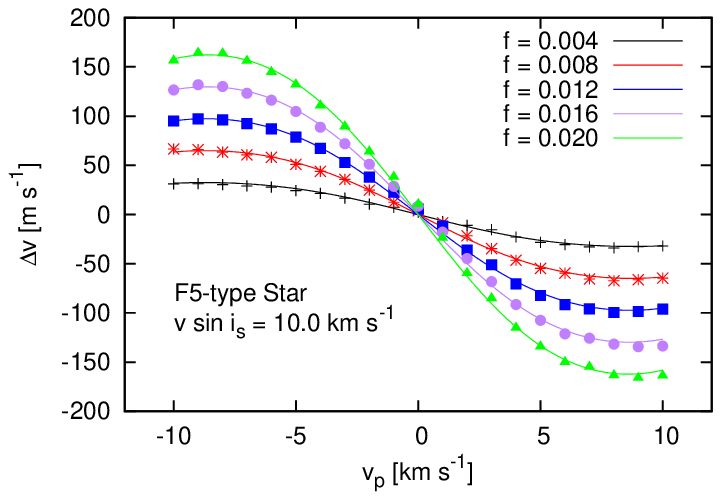} 
\includegraphics[width=9.5cm,clip]{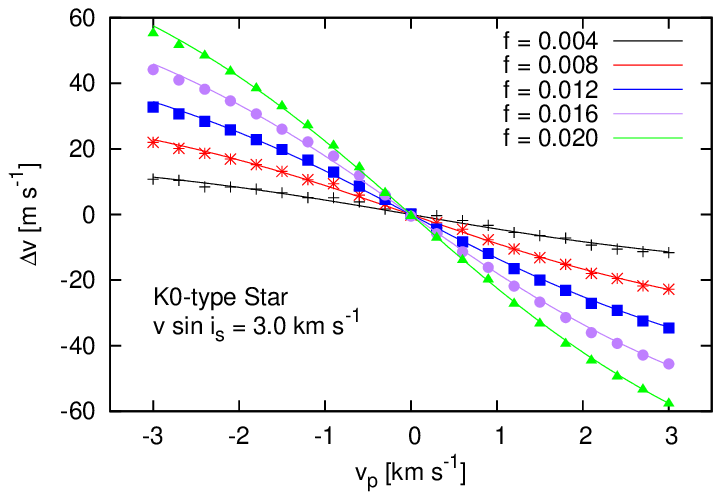} 
\caption{
The comparison between simulated results (symbols)
and Equation (\ref{eq:result}) (solid curves) for an F5 star with 
$v\sin i_s=10.0$ km s$^{-1}$ (\textit{upper})
and for a K0 star with $v\sin i_s=3.0$ km s$^{-1}$ (\textit{lower}),
respectively.
}\label{fig:F5mock}
\end{center}
\end{figure}
%%%%%%%%%%%%%%%%%%%

\item[K0 star] 
We also make the mock transit spectra for a K0 star ($T_\mathrm{eff}=5000$ K)
and put them into the RV routine. 
Since most of the K-type dwarfs are relatively slow rotators,
we adopt small rotational velocities: $v\sin i_s=1.5$, 3.0, and 4.5~km~s$^{-1}$. 
We fix the limb-darkening parameters and the macroturbulent velocity
parameter to be
$u_1=0.65$, $u_2=0.14$ \citep{Claret2004}, and $\zeta=2.8$~km~s$^{-1}$ \citep{Valenti2005}.
We compare the simulated velocity anomalies with Equation (\ref{eq:result}).
As an example we show the result for $v\sin i_s=3.0$ km s$^{-1}$ in
the lower panel of Figure \ref{fig:F5mock}, 
for which we assume $\beta=3.9$ km s$^{-1}$ and $\gamma=1.1$ km s$^{-1}$,
as estimated by analyzing the synthetic line profiles for G0 and F5 stars. 
\end{description}
All the numerical simulations show that our new analytic formula reproduces 
the simulated results within the current RV precisions.
The agreement also suggests that
we can validate the previously reported empirical relations for Subaru/HDS
\citep[i.e.,][]{Narita2009a, Narita2009b, Narita2010, Narita2010b, Hirano2011},
which are based on the similar mock data simulations.

%%%%%%%%%%%%%%%%%%%%%%%%%%%%%%%%%%%%%%%%%%%%%
\subsection{Sensitivity of the Formula to Line
  Parameters\label{sensitivity}}

In the previous subsection, we have shown that the analytic formula
(Eq.~[\ref{eq:result}]) gives an accurate description of the simulated
velocity anomalies during a transit as long as we adopt appropriate
values of $\beta$ and $\gamma$ for a given type of star.  However, there
are several practical issues that must be addressed before the
analytic formula is applied to real data analysis. First, the
macroturbulent velocity parameters that we assumed in both the
application of the analytic formula and in the mock data simulations
will not be known {\it a priori} for real stars.  Second, we have used
theoretically synthesized spectra to generate mock transit spectra,
but actual intrinsic line profiles in observed spectra may differ from
theoretical ones. Finally, although we have adopted representative
instrumental profiles in both of the analytic formula and simulation,
the instrumental profile is generally dependent on many factors such
as temperature variations, the position on CCD, etc, even if we adopt
the same spectrograph setups for observations.  Therefore, it is
important to investigate the sensitivity of Equation (\ref{eq:result})
to those parameters affecting the line profile (including the
instrumental profile).  In this subsection, we perturb the values of
those line parameters and examine the resulting changes to the outputs
of the analytic formula.

\subsubsection{Dependence on Macroturbulence\label{sec:macroeffect}}

\begin{figure}[H]
\begin{center}
\vspace{-1.8cm}
\includegraphics[width=8.2cm,clip]{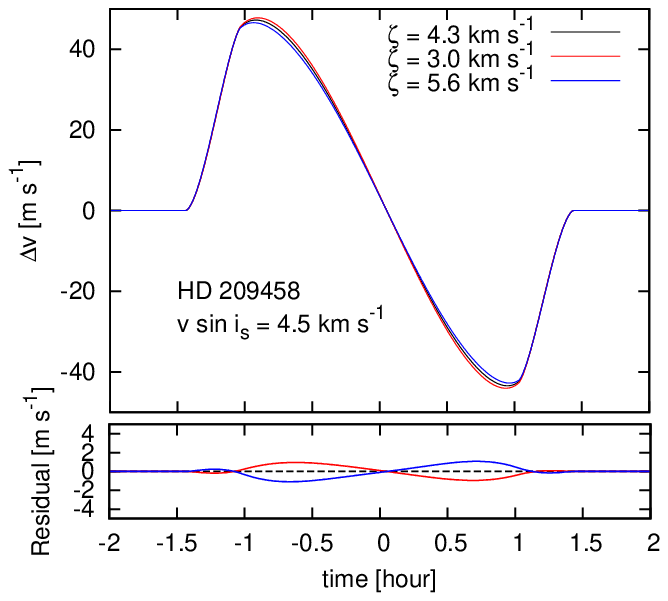} 
\includegraphics[width=8.2cm,clip]{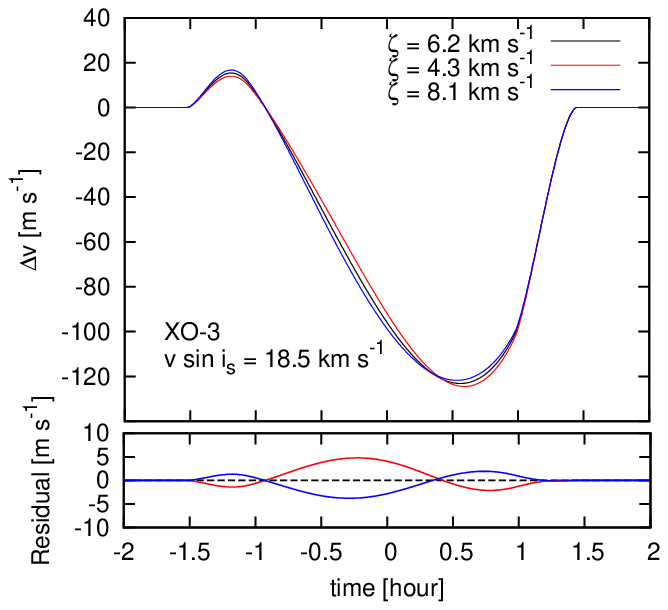} 
\includegraphics[width=8.2cm,clip]{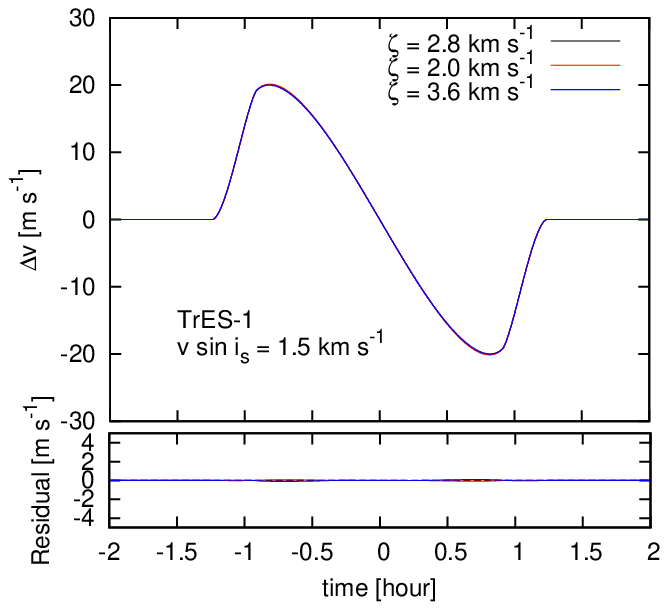} 
\vspace{-0.5cm}
\caption{Variations of RM velocity anomaly curves for 
HD~209458 (top), XO-3 (middle), and TrES-1 (bottom)
based on Equation (\ref{eq:result}).
In each panel, we change the macroturbulence parameter $\zeta$
by $\pm 30\%$ (red and blue curves) from the literature-based values
(black curves). The residuals of the red and blue curves from the black
curves are also shown at the bottom for each system.
For the spin-orbit misalignment angle $\la$, we assumed
$\la=0^\circ$ for HD 209458 and TrES-1, and 
$\la=37.3^\circ$ \citep{Winn2009}, respectively.
}\label{fig:zeta1}
\end{center}
\end{figure}

First we consider changes in the macroturbulence parameter. The
macroturbulence dispersion $\zeta$ for a G0-type star like HD~209458
is empirically estimated as $\zeta=4.3$ km s$^{-1}$
\citep{Valenti2005}.  We here change it by $\pm 30\%$ with respect to
that value (i.e.\ $\zeta=3.0$ and 5.6 km s$^{-1}$) and plot the
velocity anomaly curves for HD~209458 assuming the other line
parameters as $(\beta,\gamma) =(4.0$ km s$^{-1}$,0.9 km s$^{-1}$) and
the stellar rotational velocity of $v\sin i_s=4.5$ km s$^{-1}$.  The
top panel in Figure \ref{fig:zeta1} shows the three cases of the
macroturbulence dispersion: $\zeta=4.3$ km s$^{-1}$(black), 3.0 km
s$^{-1}$(red), and 5.6 km s$^{-1}$ (blue).

%%%%%%%%%%%%%%%%%%%
%%%%%%%%%%%%%%%%%%%
Inspection of Figure \ref{fig:zeta1} shows that the RV difference
between the curves is at most $\sim 1$ m s$^{-1}$ ($\sim 1.5\%$ of the
representative velocity anomaly $(R_p/R_s)^2 v\sin i_s$ for
HD~209458), which is less than the RV precision from an observational
point of view.  Therefore, the choice of $\zeta$ for this type of star
does not significantly affect the results.

We also show the comparisons for other two systems: XO-3 and TrES-1.
For an F5 star XO-3, we adopt $\beta=4.0$ km
s$^{-1}$, $\gamma=0.9$ km s$^{-1}$, and $v\sin i_s=18.5$ km s$^{-1}$,
and try three different cases of $\zeta$: 4.3, 6.2, and 8.1 km
s$^{-1}$ (changed by $\pm30\%$ around the central value).  Likewise,
we compute the RM velocity anomaly for TrES-1, assuming $\beta=3.9$ km
s$^{-1}$, $\gamma=1.1$ km s$^{-1}$ and $v\sin i_s=1.5$ km s$^{-1}$. We
choose values for the macroturbulence dispersion of $\zeta=2.0$, 2.8,
and 3.6 km s$^{-1}$.  The results are shown in the middle and bottom
panels in Figure \ref{fig:zeta1}.  The RV differences of the colored
curves from the black ones are at most $\approx$ 5 m s$^{-1}$ ($\sim
3.5\%$ of $(R_p/R_s)^2 v\sin i_s$) for XO-3 and $\sim$0.1 m s$^{-1}$
($\sim 0.4\%$ of $(R_p/R_s)^2 v\sin i_s$) for TrES-1,
respectively. These values are less than the RV
precision that has been achieved for each system, which are
approximately 8~m~s$^{-1}$ \citep{Winn2009} and 10~m~s$^{-1}$ \citep{Narita2007}, 
respectively.

\subsubsection{Dependence on $\beta$ and $\gamma$}

\begin{figure}[H]
\begin{center}
\includegraphics[width=8.2cm,clip]{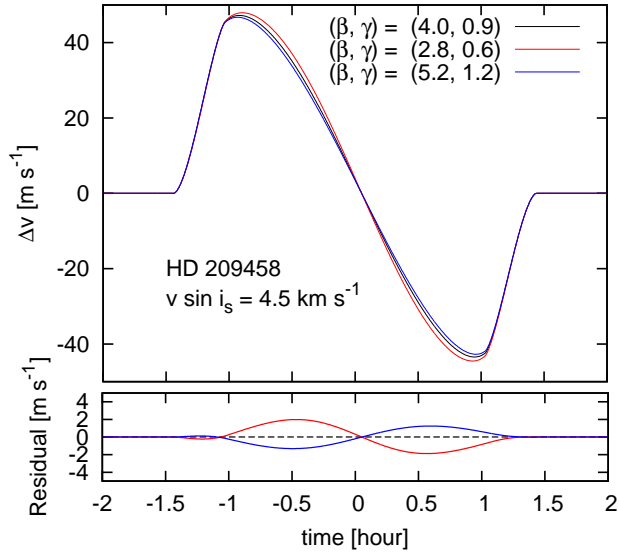} 
\caption{
The variation of RM velocity anomaly curve for 
HD~209458.
We change the Gaussian and Lorentzian dispersion $(\beta,\gamma)$ 
in Equation (\ref{eq:result})
by $\pm 30\%$ (red and blue curve) from the values
based on the line analysis (black curve). 
The residuals of the red and blue curves from the black
one are also shown at the bottom.
}\label{fig:bg}
\end{center}
\end{figure}

Next, we focus on the Gaussian and Lorentzian components of line
profiles.  Fixing the values of macroturbulence parameter $\zeta$, we
change $\beta$ and $\gamma$ in Equation (\ref{eq:result}), and plot the RM
curves for the three systems above.  Figure \ref{fig:bg} presents the
three different cases of $(\beta,\gamma)$ for HD~209458.  We change
both of $\beta$ and $\gamma$ by $\pm30\%$ from the values estimated by
the spectral line analysis (Appendix \ref{sec:app2.5}) while we fix
$\zeta$ at 4.3 km s$^{-1}$.  According to Figure \ref{fig:bg}, the
largest discrepancy of the two colored curves from the black one is
$\lesssim2$ m s$^{-1}$ ($\sim 3.1\%$ of $(R_p/R_s)^2 v\sin i_s$).
Although this is still comparable to or less than the RV precision for
this type of stars, it is slightly larger than the difference in the
curves for different $\zeta$ (see the top panel of Figure
\ref{fig:zeta1}).

%%%%%%%%%%%%%%%%%%%
%%%%%%%%%%%%%%%%%%%
We also compare the analytic RM curves for different values of $(\beta,\gamma)$
in case of XO-3 (F5-type star) and TrES-1 (K0-type star).
As a result, the differences in the RM curves are at most $\lesssim 3$ m s$^{-1}$
($\sim 2.1\%$ of $(R_p/R_s)^2 v\sin i_s$)
for XO-3 and $\lesssim 0.5$ m s$^{-1}$ for TrES-1
($\sim 2.0\%$ of $(R_p/R_s)^2 v\sin i_s$). 
These deviations are less than the current RV precision for each type
of star.
For a rapidly rotating star such as XO-3, the line profile is mainly determined 
by the stellar rotation and macroturbulence, and thus the intrinsic line profiles
(thermal and Lorentzian broadening) are likely to be less important. 
On the other hand, for slowly rotating late-type
stars as TrES-1, the intrinsic line parameters $(\beta,\gamma)$ are
important for describing the line shapes. 
However, since the RM velocity anomaly is well approximated
as $\D v\approx -fv_p$ for narrow line profiles (slow rotators) and 
the velocity amplitude due to the RM effect is comparably small, an
inaccurate estimation for $(\beta,\gamma)$ or a variation of the instrumental 
profiles (less than $\sim 30\%$) 
are not important for slowly rotating stars.

In summary, the relative importance of the various line parameters
depends on the stellar rotational velocity and the stellar type.  For
rapidly rotating stars ($v\sin i_s\gtrsim 10$ km s$^{-1}$), the
macroturbulence dispersion is more important than thermal and natural
profiles, while the Gaussian and Lorentzian dispersions become more
significant as for moderately rotating stars (3.0 km s$^{-1}\lesssim
v\sin i_s\lesssim 10$ km s$^{-1}$). Neither of the effects is
important for slowly rotating stars ($v\sin i_s\lesssim 3.0$ km
s$^{-1}$).

%%%%%%%%%%%%%%%%%%%%%%%%%%%%%%%%%%%%%%%%%%%%%
\subsection{Comparison with the Published ``Calibrations'' for Keck/HIRES}

In Section \ref{sec:tests}, we compared the analytic formula with
simulated results based on the analysis routine for Subaru/HDS. 
%As a result, we have found 
We have confirmed that Equation (\ref{eq:result}) well
approximates the simulated results with deviations less than
$\sim 0.5\%$ of the typical velocity anomaly scale (i.e.\ $fv\sin
i_s$).  It is very interesting to see if the new formula
(Eq.~[\ref{eq:result}]) also agrees with the simulated results in
previously published papers using some instrument other than
Subaru/HDS.  If they are shown to be consistent with each other, it
suggests we no longer need to perform numerical case-by-case mock data
simulations for RM analyses.  If they do not agree, it may justify
further investigation into which approach is more accurate.  Here, we
consider four systems for which the RM effect has been measured with
Keck/HIRES: HAT-P-4, HAT-P-14, XO-3, and HAT-P-11.  We compare
Equation~(\ref{eq:result}) with the published empirical relations
based on mock data simulations for Keck/HIRES.

As discussed in Section \ref{sec:IP}, the instrumental profile affects
the velocity anomaly due to the RM effect.  Therefore, in order to
make a comparison between Equation (\ref{eq:result}) and the empirical
formulae, we need to know the representative dispersion of the
instrumental profile of Keck/HIRES. We here adopt
$\beta_\mathrm{IP}=2.8$ km s$^{-1}$ in terms of velocity (a
representative dispersion for the spectral resolution of $R=65000$).
Since other parameters are intrinsic stellar properties, we adopt
similar values of $\beta_0$, $\gamma$, and $\zeta$ to the ones adopted
for the comparison in Section \ref{sec:tests} to draw the analytic RM
curves.

%%%%%%%%%%%%%%%%%%%
\begin{table}[H]
\caption{
Calibrations of the RM effect drawn from the literature,
and choices of the stellar line parameters used in Figure \ref{fig:keck}.
The stellar parameters $(\beta,\gamma,\zeta, v\sin i_s)$ are expressed in km s$^{-1}$.
}\label{table1}
\begin{center}
\begin{tabular}{ccccccc}
\hline\hline
System & Empirical Relation &$\beta$&$\gamma$ 
& $\zeta$& $v \sin i_s$ &Reference \\\hline
HAT-P-4 & $\D v=-f v_p[1.36-0.628\left(\frac{v_p}{v\sin i_s}\right)^2]$ & 3.2 & 0.9
& 4.1 & 5.5 & \citet{Winn2011} \\
HAT-P-14 & $\D v=-f v_p[1.58-0.883\left(\frac{v_p}{v\sin i_s}\right)^2]$ & 3.3 & 0.9
& 6.5 & 8.4 & \citet{Winn2011} \\
XO-3 & $\D v=-f v_p[1.644-1.036\left(\frac{v_p}{v\sin i_s}\right)^2]$ & 3.3 & 0.9
& 6.2 & 18.5 & \citet{Winn2009} \\
HAT-P-11 & $\D v=-f v_p$ & 3.2 & 1.1
& 2.5 & 1.5 & \citet{Winn2010} 
\\\hline
\end{tabular}
\end{center}
\end{table}
%%%%%%%%%%%%%%%%%%%
%%%%%%%%%%%%%%%%%%%
\begin{figure}[H]
\begin{center}
\begin{minipage}{0.49\hsize}
\begin{center}
\includegraphics[width=8.2cm]{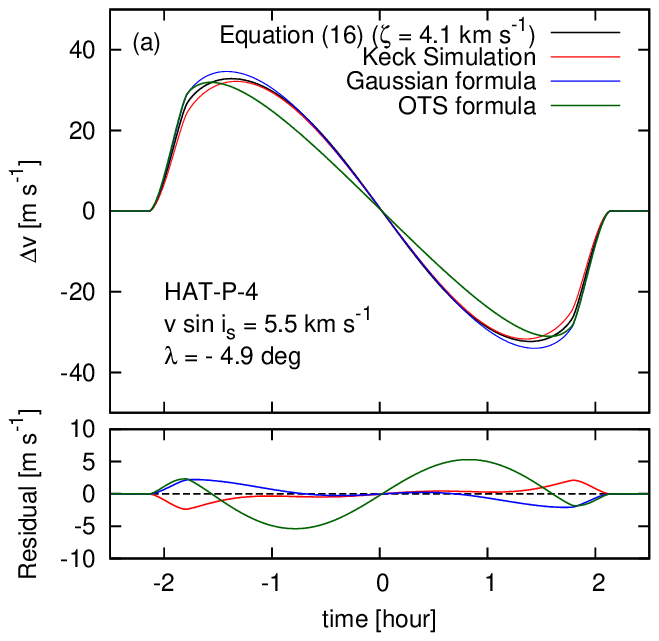}
\end{center}
\end{minipage}
\begin{minipage}{0.49\hsize}
\begin{center}
\includegraphics[width=8.2cm]{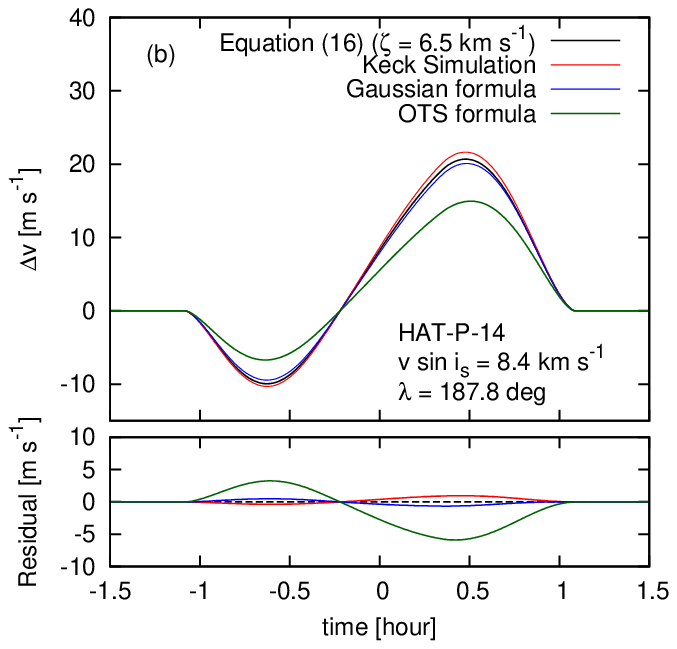}
\end{center}
\end{minipage}
\begin{minipage}{0.49\hsize}
\begin{center}
\includegraphics[width=8.0cm]{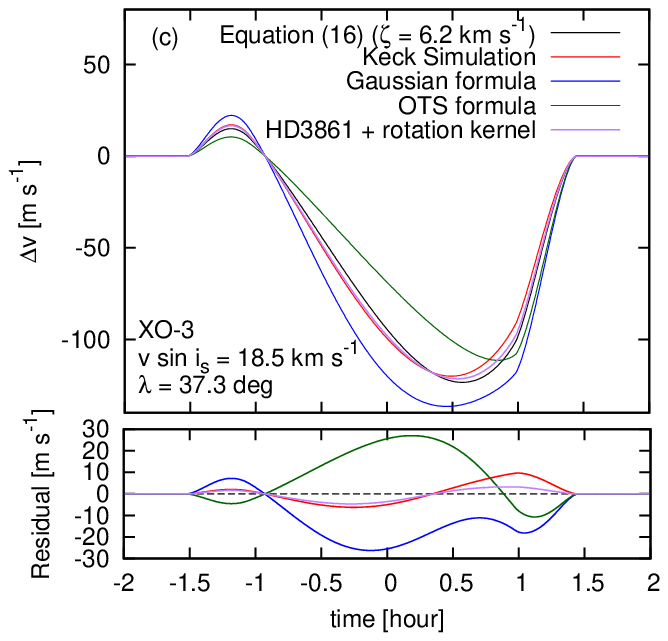}
\end{center}
\end{minipage}
\begin{minipage}{0.49\hsize}
\begin{center}
\includegraphics[width=8.2cm]{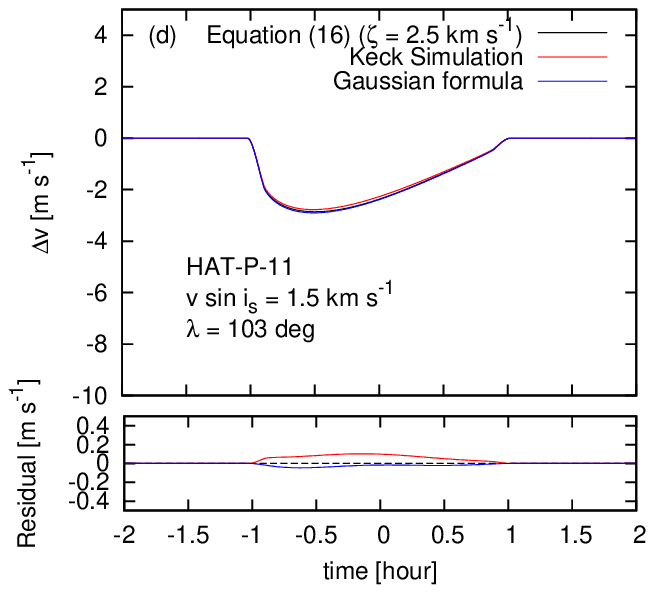}
\end{center}
\end{minipage}
\caption{The comparison between Equation (\ref{eq:result}) (black)
and the empirical relations in literatures (red) for (a) HAT-P-4,  (b) HAT-P-14,
(c) XO-3, and (d) HAT-P-11. For reference, analytic curves based on 
the Gaussian formula by \citet{Hirano2010} and 
the OTS formula are plotted in blue and green, respectively. 
The residuals of the empirical curves and the previous analytic formulae from 
Equation (\ref{eq:result}) are also shown at the bottom in each panel.
}\label{fig:keck}
\end{center}
\end{figure}
%%%%%%%%%%%%%%%%%%%

Figure \ref{fig:keck} shows the comparison between Equation
(\ref{eq:result}) and the empirical relations in literatures for the
four systems. The empirical relations in the literatures are
summarized in the second column of Table \ref{table1} and shown with
red curves in Figure \ref{fig:keck}.  Analytic curves
(Eq.~[\ref{eq:result}]) are shown in black.  The adopted stellar line
parameters to draw the analytic curves are also summarized in Table
\ref{table1}.  In Table \ref{table1}, $\beta$, $\gamma$, $\zeta$, and
$v\sin i_s$ are expressed in km s$^{-1}$.  For reference, we plot 
the Gaussian formula by \citet{Hirano2010} (i.e. Eqn.[\ref{hirano_gaussian}])
in blue and the OTS formula in green, which is expressed as $\D v = -fv_p/(1-f)$.
In drawing the Gaussian formula, we employ the width $\beta_p$
of the intrinsic spectral line (in the absence of stellar rotation)
approximated as a single Gaussian by $\beta_p=\sqrt{\beta^2+\zeta^2}$,
where $\beta$ and $\zeta$ are the values shown in Table \ref{table1}.

Regarding HAT-P-4, HAT-P-14, and HAT-P-11, for which the rotational
velocities are relatively small, Equation (\ref{eq:result}) and the
numerical calibration formulas (red) agree within a few m s$^{-1}$.  For
the HAT-P-11 system in particular, the two curves are almost
indistinguishable. This is because Equation (\ref{eq:result}) is well
approximated by $\D v\approx-fv_p$ for the case in which stellar lines are
sufficiently narrow.  On the other hand, the discrepancy between them
for the case of XO-3 is more significant ($\sim 10$ m s$^{-1}$) in
comparison with other systems although they are in much better
agreement in comparison to the previous analytic curves based on
the Gaussian and the OTS formulae.

There are at least two possible explanations for the discrepancies
between Equation (\ref{eq:result}) and the Keck simulated result.
First, the line parameters ($\beta,\gamma,\zeta$) adopted in
Equation~(\ref{eq:result}) to plot the analytic curves in Figure
\ref{fig:keck} may be significantly different from the true values for
the system. Specifically, the macroturbulence dispersion $\zeta$ is
not well known for massive stars, and different values of $\zeta$ lead
to different results as we have shown in
Section~\ref{sec:macroeffect}.  In Section \ref{sec:tests}, on the
other hand, we have compared Equation (\ref{eq:result}) with simulated
results adopting the same values of $\zeta$ in
Equation~(\ref{eq:result}) as the ones used to generate mock transit
spectra.

A second possible explanation for the discrepancy between the analytic
formula and the Keck simulation is that the coupling between the
stellar rotation and macroturbulence (which was ignored in the
numerical calibrations) has a non-negligible impact on the results.
In most cases, the mock data simulations for Keck/HIRES used real
observed spectra to create mock spectra during a transit.  For
example, to create mock spectra during a transit of the XO-3 system,
\citet{Winn2009} began with the template spectrum of HD~3861, which is
the same spectral type as XO-3 but has a smaller rotational velocity
(F5V, $v\sin i_s=2.8$ km s$^{-1}$).  Then, they broadened that
spectrum with a pure-rotational kernel to mimic the spectrum of XO-3.
This is a good choice in the sense that the starting spectrum already
involves the macroturbulence, which is not well known for those types
of stars. However, as we have emphasized, the rotational and
macroturbulence broadenings are coupled and, strictly speaking, cannot
be applied sequentially. To be more precise, we should first
deconvolve the spectrum with the macroturbulence before convolving
with a kernel including the effects of both rotation and
macroturbulence.

In order to test the second hypothesis, we perform another 
mock data simulation for XO-3. We begin with the F5-type synthetic spectrum
and broaden it with $M(v)$ in which we assume $v\sin i_s=2.8$ km s$^{-1}$
and $\zeta=6.2$ km s$^{-1}$, so that the resulting spectrum mimics
the template spectrum of HD~3861 (let us call it $T_\mathrm{HD3861}(\la)$). 
Then, we additionally broaden $T_\mathrm{HD3861}(\la)$ 
with the pure-rotational broadening kernel without macroturbulence
assuming $v\sin i_s=18.5$ km s$^{-1}$ (we call the resulting spectrum
as $T_\mathrm{XO3,Keck}(\la)$).
This process is expected to reproduce the steps taken by \citet{Winn2009}
in creating the mock spectrum of XO-3, although our starting spectrum
is a theoretically synthesized one rather than that of HD~3861.
By changing $(f,v_p)$, we generate many mock transit 
spectra from $T_\mathrm{XO3,Keck}(\la)$
and put the mock spectra into the RV analysis routine.

As a result of fitting the output velocity anomaly $\D v$ as a function
of $f$ and $v_p$, we obtain the following empirical relation:
%%%%%%%%%%%%%%%%%%%%%%%%%%%%
\begin{eqnarray}
\label{eq:xo3_empirical}
\D v= -fv_p\left[1.60 - 0.900\left(\frac{v_p}{18.5~\mathrm{km~s^{-1}}}
\right)\right]\pm \sigma_\mathrm{fit}, ~~ \sigma_\mathrm{fit}=
0.0073~\mathrm{km~s^{-1}},
\end{eqnarray}
%%%%%%%%%%%%%%%%%%%%%%%%%%%%
where $\sigma_\mathrm{fit}$ is the dispersion of the residuals of the
simulated $\D v$ from the best-fit curve.  We plot this empirical
relation in Figure \ref{fig:keck} (c) in purple.  Although we still see
differences between the Keck simulation (red) and our new
simulation (purple), they are reduced, and the new empirical relation
follows more closely the variation seen in the published numerical
calibration.  This suggests that part of the discrepancy (and perhaps
most of the discrepancy) between Equation (\ref{eq:result}) and the
Keck empirical formula is ascribed by the coupling between the stellar
rotation and macroturbulence which was neglected in
performing the numerical calibration.  Note that the RV difference
between the two simulations (red and purple) is comparable to the
dispersion of our simulated results ($\sigma_\mathrm{fit}=7.3$ m
s$^{-1}$).

%%%%%%%%%%%%%%%%%%%%%%%%%%%%%%%%%%%%%%%%%%%%%%%%%%%%%%%%%%%%%%%%%%%%%%
\section{Impact of Differential Rotations of Stars\label{sec:app3}}
One of the important applications of the new analytic formula is 
describing the stellar differential rotation in RM measurements.
\citet{Gaudi2007} previously pointed out the impact of differential rotations
via RM measurements is negligible since they are beyond the precision 
of the RV measurements.
However, their argument on the detectability is focused on relatively slow rotators.
Since the RM effect is now measured for fairly rapid rotators ($v\sin i_s\gtrsim 10$ km s$^{-1}$),
it is important to see if the impact of stellar differential rotations is still
negligible based on the precise modeling of the RM effect 
described in this paper.
In this section, we describe the impact of stellar differential rotations 
in the velocity anomaly curves.

Following \citet{Reiners2003}, we model the stellar angular velocity $\Omega$ 
as a function of the latitude $l$ on the stellar surface
($l=0$ at the stellar equator) as
%%%%%%%%%%%%%%%%%%%%%%%%%%%%
\begin{eqnarray}
\label{eq:diffrotation}
\Omega(l)=\Omega_\mathrm{eq}(1-\alpha\sin^2l),
\end{eqnarray}
%%%%%%%%%%%%%%%%%%%%%%%%%%%%
where $\Omega_\mathrm{eq}$ is the angular 
velocity at the equator. The coefficient $\alpha$ is the parameter
describing the degree of differential rotation and approximately
0.2 for the case of the Sun.
If the transiting planet is located at $(x,y)$ on the stellar disk, its latitude $l$
is estimated by
%%%%%%%%%%%%%%%%%%%%%%%%%%%%
\begin{eqnarray}
\label{eq:latitude}
\sin l=\frac{y}{R_s}\sin i_s+\sqrt{1-\frac{x^2+y^2}{R_s^2}}\cos i_s.
\end{eqnarray}
%%%%%%%%%%%%%%%%%%%%%%%%%%%%

When the host star is differentially rotating, we must replace the constant 
$\Omega$ in Equations (\ref{eq:vpdef}) and (\ref{eq:Mdxdy}) 
with Equation (\ref{eq:diffrotation}). In this case, the degeneracy between 
$\Omega$ and $\sin i_s$ can be solved; for rigidly rotating systems, 
the velocity anomaly $\D v$
depends solely on $\Omega\sin i_s$, but with a differential rotation,
the specific choice of the stellar inclination $i_s$ also affects the result.

%%%%%%%%%%%%%%%%%%%
\begin{figure}[H]
\begin{center}
\begin{minipage}{0.49\hsize}
\begin{center}
\includegraphics[width=8.2cm]{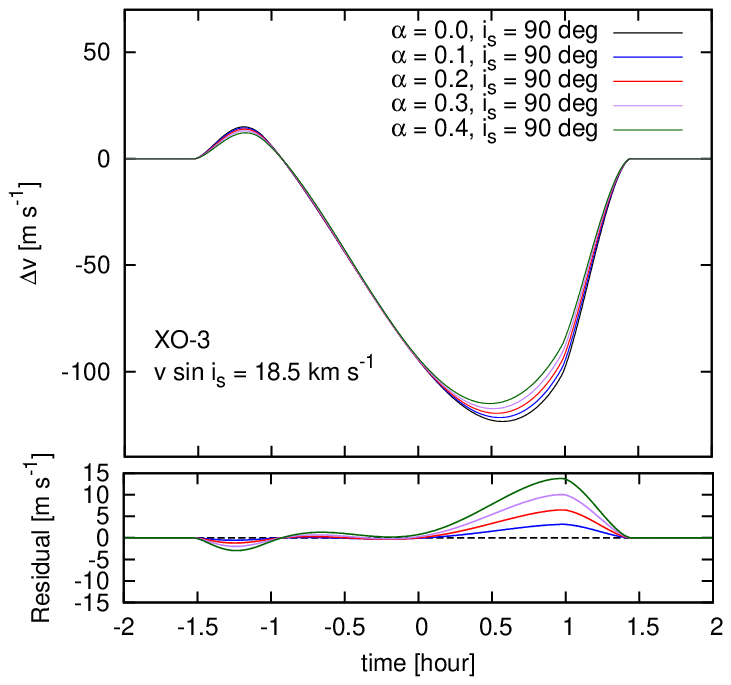}
\end{center}
\end{minipage}
\begin{minipage}{0.49\hsize}
\begin{center}
\includegraphics[width=8.2cm]{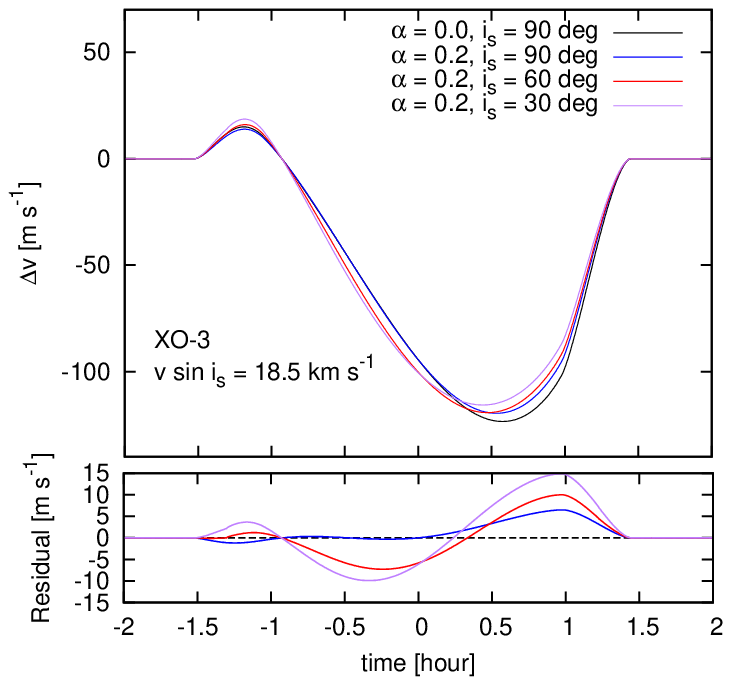}
\end{center}
\end{minipage}
\caption{
The RM velocity anomaly curves for the XO-3 system with and without 
differential rotation. The blacks lines in each panel indicate the case when the host star is
rigidly rotating while the other colors are for the cases that the star is 
differentially rotating. In the left panel, the stellar inclination is fixed 
at  $i_s=90^\circ$ while $\alpha$ is changed from 0.0
up to $0.4$.
The RM curves for various stellar inclinations are shown in the right panel along with the case
for rigid rotation.
Different colors indicate different coefficients ($\alpha$) and stellar inclinations ($i_s$)
as shown in each panel.
For each panel,
RV residuals from the rigid rotation (the black line) are shown at the bottom.
}\label{fig:diff}
\end{center}
\end{figure}
%%%%%%%%%%%%%%%%%%%
In order to evaluate the impact of differential rotations, 
we take XO-3 again as an example.
Since the XO-3 system is reported to have a large spin-orbit misalignment
and its host star has a large rotational velocity, we can expect a large impact 
of the differential rotation on RM velocity anomalies. 
Figure \ref{fig:diff} indicates the RM anomaly curves for XO-3
with and without differential rotation. 
In the left panel, we change the differential rotation coefficient $\alpha$,
fixing the stellar inclination $i_s$. We try the case of $\alpha=0.0$, 0.1, 0.2, 0.3, and 0.4.
The right panel of Figure \ref{fig:diff} indicates the variation of the RM velocity 
anomaly for the different cases of $i_s$, in which we set $i_s=90^\circ$,
$60^\circ$, and $30^\circ$ while $\alpha$ is fixed at 0.2.
In this figure, the stellar spin velocity at the equator is fixed as 
$v\sin i_s=R_s\Omega_\mathrm{eq}\sin i_s=18.5$ km s$^{-1}$.
At the bottom of each panel, we show the residual of each curve from the black 
line (rigid rotation). 

It is notable that the residuals are comparable to or rather larger than 
the RV precision for this system, which is reported to be
$\sim 8$ m s$^{-1}$ \citep{Winn2009}.
The largest impact ($\sim 15$ m s$^{-1}$) on $\D v$ occurs 
for the case of large values of $\alpha$ and small values of $i_s$, as is expected. 
%This level is within the reach of our current RV precision
%as long as the activity on the stellar surface is not so significant. 
Figure \ref{fig:diff} also shows that the stellar inclination $i_s$ significantly affects the velocity anomaly.
This result, in turn, suggests that we may be able to estimate the
stellar inclination $i_s$ from RM analyses if we can model the differential 
rotation for each type of stars and fix the coefficient $\alpha$.

%%%%%%%%%%%%%%%%%%%%%%%%%%%%%%%%%%%%%%%%%%%%%%%%%%%%%%%%%%%%%%%%%%%%%%
\section{Discussion and Summary\label{sec:summary}}

We have developed a new analytic formula (Eq.~[\ref{eq:result}]) to
describe the velocity anomaly during a planetary transit.  
We here summarize the previous findings describing the RM velocity anomaly
along with our new analytic formula:
%%%%%%%%%%%%%%%%%%%%%%%%%%%%
\begin{eqnarray}
\label{eq:summary}
\D v&=&-\frac{f}{1-f}v_p, ~~~~~\textrm{(OTS)}\\
\D v&\approx&-\left\{\frac{2\beta_\star^2}{\beta_\star^2+\beta_p^2}\right\}^{3/2}fv_p
\left[1-\frac{v_p^2}{\beta_\star^2+\beta_p^2} + 
\frac{v_p^4}{2(\beta_\star^2+\beta_p^2)^2}\right], ~~~~~
\textrm{\citep{Hirano2010}}\label{hirano_gaussian}\\
\D v&\approx& -\frac{f}{2\pi}\frac{\displaystyle \int_0^\infty 
\exp(-2\pi^2\beta^2\sigma^2-4\pi \gamma\sigma)
\tilde{M}(\sigma)\tilde{\Theta}(\sigma)\sin(2\pi \sigma v_p)\s d\sigma }
{\displaystyle\int_0^\infty \exp(-2\pi^2\beta^2\sigma^2-4\pi\gamma\sigma)
\tilde{M}(\s)\left\{\tilde{M}(\s)-f\tilde{\Theta}(\s)\cos(2\pi \s v_p)\right\}\s^2d\s},
~~\textrm{(this work)}\nonumber
\end{eqnarray}
%%%%%%%%%%%%%%%%%%%%%%%%%%%%
where $\beta_\star$ and $\beta_p$ in Equation (\ref{hirano_gaussian})
indicate the best-fit Gaussian dispersions for the stellar
template spectrum (including stellar rotation), and for the intrinsic 
line broadening in the absence of stellar rotation, respectively 
(see \citet{Hirano2010} for details).
Major improvement in our new formula is that it precisely 
modeled various effects which can possibly affect stellar line
profiles such as rotational broadening, macroturbulence, and the
instrumental profile.

Numerical simulations using the mock transit
spectra and the RV analysis routine for Subaru/HDS have shown that our
analytic formula agrees with the simulated results within $\sim 0.5\%$
of typical RM amplitudes ($\lesssim 0.25$ m s$^{-1}$ in the case of a
typical transit of a hot Jupiter).

We have also compared Equation (\ref{eq:result}) with the output of
numerical calibrations of the RM effect that have been given in the
literature, based on mock data simulations for Keck/HIRES.  We have
plotted the RM velocity anomaly curves for four existing transiting
systems: HAT-P-4, HAT-P-14, XO-3, and HAT-P-11. Equation
(\ref{eq:result}) proved to be in good agreement with previously
reported empirical relations, although some deviations were seen for
the XO-3 system.  We suggest that this disagreement comes from the
coupling between rotational and macroturbulent broadening, and perhaps
also to imperfectly known stellar line parameters. One possible means
by which this issue could be clarified is to apply our new analytic
formula to observed RVs and see if Equation (\ref{eq:result}) gives a
better fit to the data.  We do not attempt to do so in the present
paper, but we intend to reanalyze many of the existing data-sets in due
course.

%It should be emphasized that we need to know several parameters
%describing the spectral line profiles in order to apply our new formula
%to actual analyses of observed RV data. 
%Although the choice of $(\zeta,\beta,\gamma)$ does not significantly
%change the results ($\lesssim$ a few m s$^{-1}$) in most cases (\S \ref{sensitivity}), 
%it is still important to select most ``plausible" values for these parameters.
%In this paper, we have computed the Gaussian and Lorentzian dispersions $\beta$
%and $\gamma$ based on the combination of the spectral line analysis
%and the simple physical principle.

As \citet{Hirano2010} and others have pointed out, the relation
between the position of the planet and the observed RM velocity
anomaly usually does not significantly change the estimation for the
spin-orbit misalignment angle $\lambda$ since a spin-orbit
misalignment is most likely to be shown by an asymmetry in the
velocity anomaly curve during a transit.  However, when the transit
impact parameter $b$ is small, the asymmetry in the velocity anomaly
curve becomes less clear and the amplitude of a series of $\D v$ becomes
more important.

%One application of our new analytic formula (Eq.~[\ref{eq:result}]) is
%for stars with differential rotation.  As we show in
%Appendix \ref{sec:app3}, when we measure the RM effect for systems
%with differentially rotating host stars, the signal of the
%differential rotation may be detectable in some cases.  One reason why
%the detection of differential rotations in transiting systems would be
%important is that it can possibly give us some information about the
%stellar inclination $i_s$, which enables us to estimate the 3D
%structure of the spin-orbit axes.

Analytic approaches to a phenomenon give us an insight into the
behavior of the phenomenon, and sometimes reveals important aspects
which numerical approaches are likely to overlook.  We hope that our
new formula will be useful in data analyses of the RM
effect and make a contribution to the progress in this field.

%%%%%%%%%%%%%%%%%%%%%%%%%%%%%%%%%%%%%%%%%%%%%%%%%%%%%%%%%%%%%%%%%%%%%%
\acknowledgments 

We wish to acknowledge Dr. Paula R.T.\ Coelho for her kind instruction
on the use of the synthetic spectra. We are very grateful to John
Asher Johnson for helpful discussions on this topic. The data analysis
was in part carried out on common use data analysis computer system at
the Astronomy Data Center, ADC, of the National Astronomical
Observatory of Japan.  T.H.\ is 
supported by Japan Society
for Promotion of Science (JSPS) Fellowship for Research (DC1: 22-5935).
Y.S. gratefully acknowledges support from the Global
Collaborative Research Fund (GCRF) ``A World-wide Investigation of
Other Worlds'' grant and the Global Scholars Program of Princeton
University, and also from JSPS Core-to-Core Program ``International
Research Network for Dark Energy''.
J.N.W.\ acknowledges support from the NASA Origins
program (NNX11AG85G).
N.N. acknowledges a support by NINS Program for Cross-Disciplinary Study.
We would like to express special thanks the anonymous referee for the 
helpful comments and suggestions on this manuscript.

%%%%%%%%%%%%%%%%%%%%%%%%%%%%%%%%%%%%%%%%%%%%%%%%%%%%%%%%%%%%%%%%%%%%%%
\appendix

\section{Derivation of the Analytic Formula (\ref{eq:result})\label{sec:app1}}
In this appendix, we derive Equation (\ref{eq:result}) based on
Equations (\ref{eq:fstar}), (\ref{eq:transit}),  (\ref{deriva}), and (\ref{Cross}).

In order to calculate convolutions and cross-correlations, dealing with them in the 
Fourier domain, where they are expressed as products, can significantly facilitate
the computation.
First, we Fourier-transform Equations (\ref{eq:fstar}) and (\ref{eq:transit}) as
%%%%%%%%%%%%%%%%%%%%%%%%%%
\begin{eqnarray}
\ti{\F}_\mathrm{star}(\s)&=&-\ti{S}(\s)\ti{M}(\s),\\
\ti{\F}_\mathrm{transit}(\s)&=&-\ti{S}(\s)\ti{M}(\s)+f \ti{S}(\s)\ti{\Theta}(\s)e^{2\pi i\s v_p},
\end{eqnarray}
%%%%%%%%%%%%%%%%%%%%%%%%%%
where the Fourier transform of an arbitrary function $F(v)$ is defined as
%%%%%%%%%%%%%%%%%%%%%%%%%%%%
\begin{eqnarray}
\label{eq:fourier}
\tilde{F}(\sigma)\equiv\integ F(v)e^{-2\pi i \sigma v}dv,
\end{eqnarray}
%%%%%%%%%%%%%%%%%%%%%%%%%%%%
so that $\ti{M}(\s)$ and $\ti{\Theta}(\s)$ are explicitly given by Equations (\ref{Mfourier})
and (\ref{Thetafourier}), respectively (see Appendix \ref{sec:app2}).
Thus, the Fourier transform of the cross-correlation function $C(x)$ becomes
%%%%%%%%%%%%%%%%%%%%%%%%%%
\begin{eqnarray}
\label{tildeC}
\tilde{C}(\s) &=& \ti{\F}_\mathrm{star}(\s)\cdot \ti{\F}_\mathrm{transit}^*(\s)\nonumber\\
&=&\left\{\tilde{S}(\s)\right\}^2\tilde{M}(\s)\left[\tilde{M}(\s)-f\ti{\Theta}(\s)e^{-2\pi i\s v_p}\right].
\end{eqnarray}
%%%%%%%%%%%%%%%%%%%%%%%%%%
The Fourier transforms of the Gaussian and Lorentzian functions
respectively are written as
%%%%%%%%%%%%%%%%%%%%%%%%%%%%
\begin{eqnarray}
G(v)=\frac{1}{\beta\sqrt{\pi}}e^{-v^2/\beta^2}&\Longrightarrow& 
\tilde{G}(\s)=e^{-\pi^2\beta^2\s^2},\label{gaussfourier}\\
L(v)=\frac{1}{\pi}\frac{\gamma^2}{v^2+\gamma^2}&\Longrightarrow& 
\tilde{L}(\s)=e^{-2\pi \gamma |\s|}.\label{lorentzfourier}
\end{eqnarray}
%%%%%%%%%%%%%%%%%%%%%%%%%%%%
When we assume the intrinsic line profile as 
$S(v)=V(v;\beta, \gamma)=G(v:\beta)*L(v;\gamma)$
(the Voigt function), the Fourier transformation of $S(v)$ becomes
%%%%%%%%%%%%%%%%%%%%%%%%%%%%
\begin{eqnarray}
\tilde{S}(\s)=e^{-\pi^2\beta^2\s^2-2\pi \gamma |\s|}.
\end{eqnarray}
%%%%%%%%%%%%%%%%%%%%%%%%%%%%
In this case, 
the inverse Fourier transformation for $\ti{C}(\s)$ is expressed as
%%%%%%%%%%%%%%%%%%%%%%%%%%
\begin{eqnarray}
\label{CbyR}
C(x)&=& \integ \ti{C}(\s)e^{2\pi i\s x}d\s\nonumber\\
&=&\integ \exp(-2\pi^2\beta^2\s^2 - 4\pi \gamma |\s| + 2\pi i \s x)\tilde{M}(\s)\left(
\tilde{M}(\s)-f\ti{\Theta}(\s)e^{-2\pi i \s v_p}\right)d\s \nonumber\\
&=&2\int_0^\infty\exp(-2\pi^2\beta^2\s^2-4\pi \gamma\s)
\left\{\tilde{M}(\s)\right\}^2\cos(2\pi \s x) d\s 
\nonumber\\
&&-2f\int_0^\infty\exp(-2\pi^2\beta^2\s^2-4\pi \gamma\s)\tilde{M}(\s)\ti{\Theta}(\s)
\cos\left\{2\pi \s(v_p-x)\right\} d\s.
\end{eqnarray}
%%%%%%%%%%%%%%%%%%%%%%%%%%
The derivative of Equation (\ref{CbyR}) with respect to $x$ is
%%%%%%%%%%%%%%%%%%%%%%%%%%
\begin{eqnarray}
\label{dCdx}
\frac{dC(x)}{dx}&=&-4\pi\int_0^\infty\exp( -2 \pi^2\beta^2\s^2-4\pi \gamma\s)
\left\{\tilde{M}(\s)\right\}^2\s \sin(2\pi\s x) d\s \nonumber\\
&&-4\pi f\int_0^\infty \exp( -2 \pi^2\beta^2\s^2-4\pi \gamma\s)
\tilde{M}(\s)\ti{\Theta}(\s)\s \sin\{2\pi\s(v_p- x)\}d\s .
\end{eqnarray}
%%%%%%%%%%%%%%%%%%%%%%%%%%
The maximum of the cross-correlation 
function is at $x=\D v$, where
%%%%%%%%%%%%%%%%%%%%%%%%%%
\begin{eqnarray}
\label{maximum}
&&\int_0^\infty \exp( -2 \pi^2\beta^2\s^2-4\pi \gamma\s)
\left\{\tilde{M}(\s)\right\}^2\s \sin(2\pi\s \D v)d\s \nonumber\\
&&=-f\int_0^\infty \exp( -2 \pi^2\beta^2\s^2-4\pi \gamma\s)
\tilde{M}(\s)\ti{\Theta}(\s)\s \sin\{2\pi\s(v_p-\D v)\}d\s.
\end{eqnarray}
%%%%%%%%%%%%%%%%%%%%%%%%%%
Because of the exponential factors, the integrals in Equation (\ref{maximum}) 
can be cut off at $ \s \sim 1/2\beta$. 
Moreover, since the Gaussian width parameter $\beta$ is of the order of 
several km s$^{-1}$ and the velocity anomaly is typically less than one
hundred m s$^{-1}$, we obtain
%%%%%%%%%%%%%%%%%%%%%%%%%%
\begin{eqnarray}
2\pi\s\D v \lesssim 0.1,
\end{eqnarray}
%%%%%%%%%%%%%%%%%%%%%%%%%%
and we can safely approximate as
$\sin(2\pi\s \D v)\approx 2\pi\s \D v$.
Thus, using the identity for trigonometric functions:
%%%%%%%%%%%%%%%%%%%%%%%%%%
\begin{eqnarray}
\sin\{2\pi\s(v_p-\D v)\}&=&\sin(2\pi\s v_p)\cos(2\pi\s\D v)-\cos(2\pi\s v_p)
\sin(2\pi\s\D v)\nonumber\\
&\approx& \sin(2\pi \s v_p)-2\pi\s\D v\cos(2\pi \s v_p),
\end{eqnarray}
%%%%%%%%%%%%%%%%%%%%%%%%%%
%we obtain the following expression:
%%%%%%%%%%%%%%%%%%%%%%%%%%
%\begin{eqnarray}
%&&2\pi\D v \integ  \exp( -2 \pi^2\beta^2\s^2-4\pi \gamma\s)
%\s^2\tilde{M}(\s)\left\{\tilde{M}(\s)-f\tilde{\Theta}(\s)\cos(2\pi \s v_p)\right\}d\s\nonumber\\
%&&=-f\int_0^\infty \exp( -2 \pi^2\beta^2\s^2-4\pi \gamma\s)
%\tilde{M}(\s)\ti{\Theta}(\s)\s \sin(2\pi\s v_p)d\s.
%\end{eqnarray}
%%%%%%%%%%%%%%%%%%%%%%%%%%
%Thus, 
we obtain following analytic formula for $\D v$:
%%%%%%%%%%%%%%%%%%%%%%%%%%%%
\begin{eqnarray}
\label{eq:resultappendix}
\D v\approx -\frac{f}{2\pi}\frac{\displaystyle \int_0^\infty 
\exp(-2\pi^2\beta^2\sigma^2-4\pi \gamma\sigma)
\tilde{M}(\sigma)\tilde{\Theta}(\sigma)\sin(2\pi \sigma v_p)\s d\sigma }
{\displaystyle\int_0^\infty \exp(-2\pi^2\beta^2\sigma^2-4\pi\gamma\sigma)
\tilde{M}(\s)\left\{\tilde{M}(\s)-f\tilde{\Theta}(\s)\cos(2\pi \s v_p)\right\}\s^2d\s}.
\end{eqnarray}
%%%%%%%%%%%%%%%%%%%%%%%%%%%%
Note that contrary to previously reported empirical relations, 
Equation (\ref{eq:resultappendix}) implies that the velocity anomaly
due to the RM effect is not proportional to the loss of flux $f$,
but the second order of $f$ slightly affects the results.

%%%%%%%%%%%%%%%%%%%%%%%%%%%%%%%%%%%%%%%%%%%%%%%%%%%%%%%%%%%%%%%%%%%%%%
\section{The Fourier Transformation of the Rotational and Macroturbulence Kernel\label{sec:app2}}
One of the great successes in dealing with the RM effect in the Fourier domain
is that the rotational and macroturbulence broadening kernel $M(v)$ becomes
significantly simpler in the Fourier domain.
Here, we Fourier-transform $M(v)$ and derive Equation (\ref{Mfourier}).

Since the velocity component $v$ appears only in the macroturbulence kernel
$\Theta(v)$, we can write the Fourier transformation of $M(v)$ as
%%%%%%%%%%%%%%%%%%%%%%%%%%%%
\begin{eqnarray}
\label{eq:Mfourierdef}
\tilde{M}(\sigma)&\equiv&\integ M(v)e^{-2\pi i \sigma v}dv\nonumber\\
&\propto&\integ \Theta(v-x\Omega \sin i_s)e^{-2\pi i \s v}dv.
\end{eqnarray}
%%%%%%%%%%%%%%%%%%%%%%%%%%%%
The last expression above is expressed by a linear combination of the following
form of an integral: 
%%%%%%%%%%%%%%%%%%%%%%%%%%%%
\begin{eqnarray}
\label{eq:nf1}
\integ \frac{1}{\sqrt{\pi}\kappa}\exp
\left\{-\left(\frac{v-x\Omega \sin i_s}{\kappa}\right)^2-2\pi i\s v\right\}dv,
\end{eqnarray}
%%%%%%%%%%%%%%%%%%%%%%%%%%%%
where $\kappa$ is a constant.
The exponent in the above integrand reduces to
%%%%%%%%%%%%%%%%%%%%%%%%%%%%
\begin{eqnarray}
%&&%\exp\left\{
-\left(\frac{v-x\Omega \sin i_s}{\kappa}\right)^2-2\pi i\s v
%&&=%\exp\left\{
=-\frac{(v-x\Omega\sin i_s+\pi i\kappa^2\s)^2}{\kappa^2}
+\pi \s i(\pi\kappa^2\s i-2x\Omega\sin i_s).
\end{eqnarray}
%%%%%%%%%%%%%%%%%%%%%%%%%%%%
Thus, using the Gaussian integral:
%%%%%%%%%%%%%%%%%%%%%%%%%%%%
\begin{eqnarray}
\label{eq:gauss}
\integ \frac{1}{\sqrt{\pi}\kappa}e^{-(v/\kappa)^2}dv=1,
\end{eqnarray}
%%%%%%%%%%%%%%%%%%%%%%%%%%%%
Equation (\ref{eq:nf1}) becomes 
%%%%%%%%%%%%%%%%%%%%%%%%%%%%
\begin{eqnarray}
\label{eq:nf2}
\integ \frac{1}{\sqrt{\pi}\kappa}\exp
\left\{-\left(\frac{v-x\Omega \sin i_s}{\kappa}\right)^2-2\pi i\s v\right\}dv
=\exp\{\pi i \s (\pi i \kappa^2\s-2x\Omega\sin i_s)\}.
\end{eqnarray}
%%%%%%%%%%%%%%%%%%%%%%%%%%%%
Replacing $\kappa$ with $\zeta\cos\theta$ or $\zeta\sin\theta$, 
the second line in Equation (\ref{eq:Mfourierdef}) is expressed as
%%%%%%%%%%%%%%%%%%%%%%%%%%%%
\begin{eqnarray}
\label{eq:Mfourierchange}
\integ \Theta(v-x\Omega \sin i_s)e^{-2\pi i \s v}dv
=\frac{e^{-2\pi i\s x\Omega\sin i_s}}{2}
\{e^{-(\pi\zeta\cos\theta)^2\s^2}+e^{-(\pi\zeta\sin\theta)^2\s^2}\}.
\end{eqnarray}
%%%%%%%%%%%%%%%%%%%%%%%%%%%%
Therefore, we obtain the following expression for $\tilde{M}(\s)$:
%%%%%%%%%%%%%%%%%%%%%%%%%%%%
\begin{eqnarray}
\label{eq:Mdxdy}
\tilde{M}(\s)&=&\iint_{\mathrm{entire~disk}}
\frac{1-u_1(1-\cos\theta)-u_2(1-\cos\theta)^2}{\pi(1-u_1/3-u_2/6)}\nonumber\\
&&~~~~~~~\times\frac{e^{-2\pi i\s x\Omega\sin i_s}}{2}
\{e^{-(\pi\zeta\cos\theta)^2\s^2}+e^{-(\pi\zeta\sin\theta)^2\s^2}\}
\frac{dxdy}{R_s^2}.
\end{eqnarray}
%%%%%%%%%%%%%%%%%%%%%%%%%%%%
Since the imaginary part of the integrand in Equation (\ref{eq:Mdxdy})
is an odd function with regard to $x$, it vanishes after integrated over $x$.
We then change variables as $x=r\cos\phi$ and $y=r\sin\phi$, 
which leads to 
%%%%%%%%%%%%%%%%%%%%%%%%%%%%
\begin{eqnarray}
\label{eq:Mdrdphi}
\tilde{M}(\s)&=&\int_0^{R_s}\int_0^{2\pi}
\frac{1-u_1(1-\cos\theta)-u_2(1-\cos\theta)^2}{\pi(1-u_1/3-u_2/6)}\nonumber\\
&&~~~~~~~\times\cos(2\pi \s r\Omega\sin i_s\cos\phi)
\frac{e^{-(\pi\zeta\cos\theta)^2\s^2}+e^{-(\pi\zeta\sin\theta)^2\s^2}}{2}
\frac{rdrd\phi}{R_s^2},
\end{eqnarray}
%%%%%%%%%%%%%%%%%%%%%%%%%%%%
where
%%%%%%%%%%%%%%%%%%%%%%%%%%%%
\begin{eqnarray}
\label{eq:cossin}
\cos\theta=\sqrt{1-\frac{r^2}{R_s^2}}, ~ \sin \theta=\frac{r}{R_s}.
\end{eqnarray}
%%%%%%%%%%%%%%%%%%%%%%%%%%%%
Using the following formula
%%%%%%%%%%%%%%%%%%%%%%%%%%%%
\begin{eqnarray}
\label{eq:bessel0}
\int_0^{2\pi}\cos(2\pi \s r\Omega\sin i_s\cos\phi)d\phi=2\pi J_0(2\pi \s r\Omega\sin i_s),
\end{eqnarray}
%%%%%%%%%%%%%%%%%%%%%%%%%%%%
we obtain
%%%%%%%%%%%%%%%%%%%%%%%%%%%%
\begin{eqnarray}
\label{eq:Mdr}
\tilde{M}(\s)&=&\int_0^{R_s}
\frac{1-u_1(1-\cos\theta)-u_2(1-\cos\theta)^2}{1-u_1/3-u_2/6}\nonumber\\
&&~~~~~~~%\cos(2\pi \s r\Omega\sin i_s)
\times\{e^{-(\pi\zeta\cos\theta)^2\s^2}+e^{-(\pi\zeta\sin\theta)^2\s^2}\}J_0(2\pi \s r\Omega\sin i_s)
\frac{rdr}{R_s^2},
\end{eqnarray}
%%%%%%%%%%%%%%%%%%%%%%%%%%%%
or equivalently,
%%%%%%%%%%%%%%%%%%%%%%%%%
\begin{eqnarray}
\tilde{M}(\sigma)&=&\int_0^1\frac{1-u_1(1-\sqrt{1-t^2})-u_2(1-\sqrt{1-t^2})^2}{1-u_1/3-u_2/6}\nonumber\\
&&~~~~~~~\times\{e^{-\pi^2\zeta^2\sigma^2(1-t^2)}+e^{-\pi^2\zeta^2\sigma^2 t^2}\}
J_0(2\pi\sigma v\sin i_s t)
tdt. \label{Mfourierappendix}
\end{eqnarray}
%%%%%%%%%%%%%%%%%%%%%%%%%

%%%%%%%%%%%%%%%%%%%%%%%%%%%%%%%%%%%%%%%%%%%%%%%%%%%%%%%%%%%%%%%%%%%%%%
\section{Estimation of the Effective Line Profile\label{sec:app2.5}}
Here, we describe how to estimate the ``effective" line profile
from a spectrum. First, we model the auto-correlation function of a spectral
assuming a simple analytic form of it,
and then compute the actual auto-correlation function for
the synthetic spectra that we use in the mock data simulation.

We assume that the synthetic spectrum \citep{Coelho2005} 
is expressed by a summation of 
Voigt functions (defined by Eq. [\ref{voigt}]) as
%%%%%%%%%%%%%%%%%%%%%%%%%%%%
\begin{eqnarray}
\label{spectrum_model}
T(\la)= \sum_{i}^\mathrm{\#~of~lines}k_iV(\la-d_i;\beta_i, \gamma_i),
\end{eqnarray}
%%%%%%%%%%%%%%%%%%%%%%%%%%%%
where $k_i$, $d_i$ are the equivalent width and the central wavelength
of the spectral line $i$. The Gaussian and Lorentzian dispersions of the line $i$
are also noted by $\beta_i$ and $\gamma_i$, respectively.
For simplicity, we subtract the continuum (normalized to unity) 
of the spectrum and flip the absorption lines so that the flux takes positive values.
The auto-correlation function $A(x)$ of the spectrum is defined as
%%%%%%%%%%%%%%%%%%%%%%%%%%%%
\begin{eqnarray}
A(x)&\equiv& \integ T(\la)T(\la-x)d\la\nonumber\\
&=&\integ \sum_{i}^\mathrm{\#~of~lines}
\sum_{j}^\mathrm{\#~of~lines}k_ik_j
V(\la-d_i;\beta_i, \gamma_i)
V(\la-d_j-x;\beta_j, \gamma_j)d\la.\label{A_def}
\end{eqnarray}
%%%%%%%%%%%%%%%%%%%%%%%%%%%%
Since the convolution between two Voigt functions yields another 
Voigt function with a larger dispersion, the integral above reduces to
%%%%%%%%%%%%%%%%%%%%%%%%%%%%
\begin{eqnarray}
A(x)&=& 
\sum_{i}^\mathrm{\#~of~lines}
\sum_{j}^\mathrm{\#~of~lines}k_ik_j
V(x-d_i+d_j;\sqrt{\beta_i^2+\beta_j^2}, \gamma_i+\gamma_j)\nonumber\\
&=&
\sum_{i}^\mathrm{\#~of~lines}
k_i^2V(x;\sqrt{2}\beta_i, 2\gamma_i)
+\sum_{i}^\mathrm{\#~of~lines}
\sum_{i>j}k_ik_j
V(x-|d_i-d_j|;\sqrt{\beta_i^2+\beta_j^2}, \gamma_i+\gamma_j)\nonumber\\
&&~+\sum_{i}^\mathrm{\#~of~lines}
\sum_{i>j}k_ik_j
V(x+|d_i-d_j|;\sqrt{\beta_i^2+\beta_j^2}, \gamma_i+\gamma_j).\label{A_x}
\end{eqnarray}
%%%%%%%%%%%%%%%%%%%%%%%%%%%%
The first term in the last expression indicates the sum of all the absorption lines
broadened by factors of $\sqrt{2}$ for Gaussian and 2 for Lorentzian profiles.
Here, we use this term to calculate the ``effective" line profile of all the spectral lines;
we define the effective Gaussian and Lorentzian dispersions $\bar{\beta}$ and
$\bar{\gamma}$ as 
%%%%%%%%%%%%%%%%%%%%%%%%%%%%
\begin{eqnarray}
\label{effective_profile}
\sum_{i}^\mathrm{\#~of~lines}
k_i^2V(x;\sqrt{2}\bar{\beta}, 2\bar{\gamma})
\equiv 
\sum_{i}^\mathrm{\#~of~lines}
k_i^2V(x;\sqrt{2}\beta_i, 2\gamma_i).
\end{eqnarray}
%%%%%%%%%%%%%%%%%%%%%%%%%%%%
Although a summation of Voigt functions is not necessarily 
expressed by a Voigt function, this treatment above is justified as long as
all the values of $\beta_i$ and $\gamma_i$ do not differ significantly 
from line to line.

The second and third terms in Equation 
(\ref{A_x}) mean the correlations between the lines
separated by $|d_i-d_j|$. It is plausible to assume that this separation $|d_i-d_j|$
is distributed randomly in a spectrum. In this case, the second and third terms become
the summations of a bunch of Voigt functions centered at randomly placed $x$.
Assuming the number of spectral lines is large enough ($i\gtrsim 1000$), 
the second and third terms altogether becomes almost constant as a function of $x$
\footnote{In reality, since we deal with a limited wavelength range of the spectrum,
the auto-correlation function $A(x)$ slowly decreases as a function of $|x|$.}. 
Therefore, we can approximate Equation (\ref{A_x}) around $x=0$ as
%%%%%%%%%%%%%%%%%%%%%%%%%%%%
\begin{eqnarray}
\label{A_approx}
A(x)\approx \bar{k} V(x;\sqrt{2}\bar{\beta}, 2\bar{\gamma}) + C, ~~~~\bar{k}\equiv \sum_{i}^\mathrm{\#~of~lines}
k_i^2,
\end{eqnarray}
%%%%%%%%%%%%%%%%%%%%%%%%%%%%
where $C$ is the constant originated from the second and third terms in Equation 
(\ref{A_x}).

Next, we try to compute the auto-correlation function $A(x)$ for the 
actual synthetic spectra we use to generate the mock transit spectra.
We take the G0-type spectrum described in Section \ref{s:simulation},
subtract the continuum, and flip the lines.  
Then, we compute the auto-correlation function $A(x)$ defined as
Equation (\ref{A_def}). Note that we cut the spectrum so that it 
ranges from $\sim 5000~\mathrm{\AA}$ to $\sim 6000~\mathrm{\AA}$, where 
precise radial velocity analyses are usually performed. 

%%%%%%%%%%%%%%%%%%%
\begin{figure}[H]
\begin{center}
\includegraphics[width=9cm,clip]{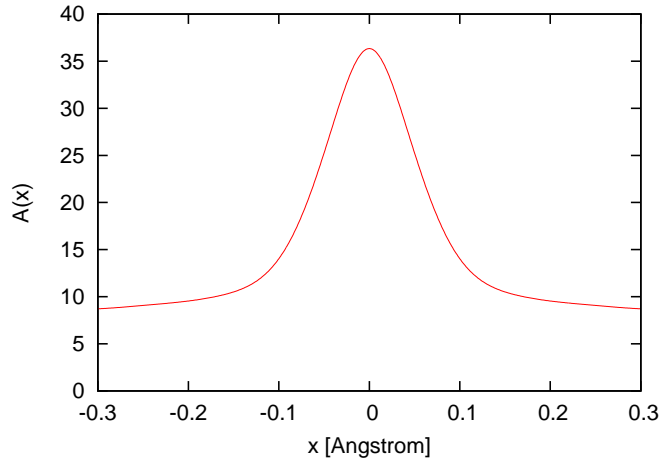} 
\caption{The auto-correlation function $A(x)$ of the synthetic spectrum 
for a G0-type star, defined by Equation (\ref{A_def}). 
The wavelength range used in the calculation is $5000~\mathrm{\AA}\lesssim 
\la\lesssim 6000~\mathrm{\AA}$.
}\label{fig:autocorrelation}
\end{center}
\end{figure}
%%%%%%%%%%%%%%%%%%%
Figure \ref{fig:autocorrelation} shows the resultant auto-correlation function $A(x)$
of the synthetic spectrum for a G0-type star used in the mock data simulation around $x=0$.
As is expected from the above theoretical prediction, it has a strong 
Voigt-shaped peak at $x=0$
with an offset due to the correlations with neighboring lines. 
We fit this auto-correlation function assuming a form of Equation (\ref{A_approx}).
Since the Gaussian and Lorentzian dispersion $\beta$ and $\gamma$ are 
strongly correlated, we fix $\bar{\beta}$ as $\bar{\beta}=\beta_0=1.7$ km s$^{-1}$, 
which is calculated by Equation (\ref{eq:beta0}), and let only $\bar{k}$, $\bar{\gamma}$, 
and $C$ be free.
As a result, we find the best-fit value of $\bar{\gamma}$ as 
$\bar{\gamma}\sim 0.94$ km s$^{-1}$. In this paper, we use these values 
in comparing the analytic formula with the simulated results for a G0-type star.

%%%%%%%%%%%%%%%%%%%%%%%%%%%%%%%%%%%%%%%%%%%%%%%%%%%%%%%%%%%%%%%%%%%%%%

%%%%%%%%%%%%%%%%%%%%%%%%%%%%%%%%%%%%%%%%%%%%%%%%%%%%%%%%%%%%%%%%%%%%%%

\end{document}